\documentclass[fleqn,usenatbib]{mnras}

\usepackage{newtxtext,newtxmath}

\usepackage[T1]{fontenc}

\DeclareRobustCommand{\VAN}[3]{#2}
\let\VANthebibliography\thebibliography
\def\thebibliography{\DeclareRobustCommand{\VAN}[3]{##3}\VANthebibliography}

\usepackage{graphicx}	
\usepackage{caption}
\usepackage{subcaption}
\usepackage{amsmath}	
\usepackage{comment}
\usepackage{wasysym} 
\usepackage{ulem}
\usepackage{xspace} 
\usepackage{multirow,multicol}
\usepackage{xcolor}
\usepackage{dblfloatfix}

\newcommand{\teff}{$T\rm_{eff}$\xspace}

\newcommand{\logg}{$\rm{\log}g$\xspace}
\newcommand{\taueff}{$\tau\rm_{ACF}$\xspace}
\newcommand{\numax}{$\nu\rm_{max}$\xspace}
\newcommand{\rms}{$\sigma$\xspace}
\newcommand\avg[1]{\ensuremath{\left\langle #1 \right\rangle}\xspace}
\defcitealias{Kallinger2014}{K14}

\title[Scaling relations of convective granulation noise]{Scaling relations of convective granulation noise across the HR diagram from 3D stellar atmosphere models} %

\author[L. F. Rodr\'iguez D\'iaz et al.]{
Luisa Fernanda Rodr\'iguez D\'iaz,$^{1}$\thanks{E-mail: luisa.rodriguez@phys.au.dk (LFRD)}
Lionel Bigot$^{2}$,
V\'ictor Aguirre Børsen-Koch$^{1}$,
Mikkel~N.~Lund$^{1}$,\and
Jakob Lysgaard Rørsted$^{1}$,
Thomas Kallinger$^{3}$,
Sophia Sulis$^{4}$,
David Mary$^{2}$
\\
$^{1}$Stellar Astrophysics Centre, Department of Physics and Astronomy, Aarhus University, Ny Munkegade 120, DK-8000 Aarhus C, Denmark\\
$^{2}$Universit\'e C\^ote d'Azur, Observatoire de la C\^ote d'Azur, CNRS, Lagrange UMR 7293, CS 34229, 
 06304, Nice Cedex 4, France\\
$^{3}$Institute for Astrophysics (IfA), University of Vienna, Türkenschanzstrasse 17, 1180 Vienna, Austria\\
$^{4}$Universit\'e Aix Marseille, CNRS, CNES, LAM, Marseille, France} 

\date{Accepted XXX. Received YYY; in original form ZZZ}

\pubyear{2020}

\begin{document}

\label{firstpage}
\pagerange{\pageref{firstpage}--\pageref{lastpage}}
\maketitle

\begin{abstract}
High-precision photometric data from space missions have improved our understanding of stellar granulation. These observations have shown with precision the stochastic brightness fluctuations of stars across the HR diagram, allowing us to better understand how stellar surface convection reacts to a change in stellar parameters. These fluctuations need to be understood and quantified in order to improve the detection and characterization of exoplanets. In this work, we provide new scaling relations of two characteristic properties of the brightness fluctuations time series, the standard deviation (\rms) and the auto-correlation time (\taueff). This was done by using long time series of 3D stellar atmosphere models at different metallicities and across the HR diagram, generated with a 3D radiative hydrodynamical code: the STAGGER code. We compared our synthetic granulation properties with the values of a large sample of \textit{Kepler} stars, and analyzed selected stars with accurate stellar parameters from the \textit{Kepler} LEGACY sample. Our 3D models showed that \rms$\propto\nu\rm_{max}^{-0.567\pm0.012}$ and \taueff$\propto\nu\rm_{max}^{-0.997\pm0.018}$  for stars at solar metallicity. We showed that both \rms and \taueff decrease with metallicity, although the metallicity dependence is more significant on \rms. Unlike previous studies, we found very good agreement between \rms from \textit{Kepler} targets and the 3D models at \logg$\leq3.5$, and a good correlation between the stars and models with \logg$\geq3.5$. For \taueff, we found that the 3D models reproduced well the \textit{Kepler} LEGACY star values. Overall, this study shows that 3D stellar atmosphere models reproduce the granulation properties of stars across the HR diagram.
\end{abstract}




\section{Introduction}
Thanks to photometric space missions such as \textit{CoRoT} \citep{Baglin2006}, \textit{Kepler} \citep{Borucki2010} and K2 \citep{Howell2014}
our knowledge about exoplanets orbiting distant stars has vastly increased, while also contributing to improving our understanding of stellar atmospheres. Particularly, these missions have driven the efforts of the last couple of years to model stellar granulation and stellar activity \citep[see e.g.][]{Pereira2019, Barros2020}, with the aim to detect and characterize small exoplanets with TESS \citep{Ricker2015} and PLATO \citep{Rauer2014}. 

Many attempts have been made to characterize the stellar signal, both from theoretical and observational perspectives. From the theoretical approach, granulation noise properties have been studied using 3D radiation-hydrodynamic simulations of stellar atmospheres. Granulation is a time-dependent phenomenon associated with heat transport by convection on horizontal scales, due to radiative cooling \citep{Nordlund2009}. As a result, the stellar surface is filled with bright, hot granules and dark, cooler intergranular lanes. This interplay of hot up- and cool down-flows is also responsible for brightness and temperature fluctuations over time on stellar surfaces. This is defined as stellar noise. 

\cite{Trampedach1998, Ludwig2006} proposed a formalism to use 3D stellar atmosphere models to derive disk-integrated brightness fluctuations, as seen by an observer. This formalism was implemented in \cite{Mathur2011}, where they used 3D models at solar metallicity from \cite{Trampedach2013}, to evaluate the granulation-induced fluctuations on the stellar brightness. They compared the results from the models with approximately 1000 red giant stars observed by \textit{Kepler}, which they analyzed with six different techniques to fit their power spectra. They derived relations of granulation properties, namely power and timescale, in terms of the frequency of maximum power of the p-mode spectrum \numax, and showed that the 3D models could recover the same behavior that observations showed, but did not reproduce them accurately. \cite{Tremblay2013} used 148 3D model atmospheres from the CIFIST grid \citep{Ludwig2009b} to study their granulation properties. They found that the relative intensity and characteristic timescale (\taueff) are correlated with the Mach number, and to some extent with the Péclet number and convective efficiency. Similarly, \cite{Samadi2013} analyzed the power spectra of solar-metallicity 3D models from the CIFIST grid and determined their root-mean-square (rms) brightness fluctuations and \taueff. Their results showed some scatter, especially for main-sequence (MS) models, when compared with \numax, i.e. an absence of a simple power law dependence $\nu_{\rm max}^\mathbb{\alpha}$. However, they found that the scatter decreases if the granulation properties are related to both \numax and the Mach number, with some combination of the two parameters. They also analyzed a small sample of stars observed by \textit{Kepler}, but the dependency on the Mach number could not be confirmed. Additionally, \cite{Ludwig2016} used 3D stellar atmosphere models from the CIFIST grid with metallicities equal to [Fe/H] = 0.0 and -2.0 to study the metallicity effect on the the brightness fluctuations.  Even though their \rms also increases with decreasing \numax, their data are scattered for both metallicities, similarly as the data presented in \cite{Samadi2013}. They also claimed that they could not simultaneously fit the data at both metallicities, and neither could reproduce the \rms values from \citet[hereafter \citetalias{Kallinger2014}]{Kallinger2014}, at least using the 3D solar metallicity models.

From the observational perspective, \cite{Bastien2013} found a clear correlation between brightness variations and surface gravity in Kepler light curves. This was done by isolating the variations related to granulation, corresponding to timescales shorter than 8 hours but longer than 30 minutes, called 8-hour flicker ($F_8$) \citep[see][for a more detailed description of the method]{Bastien2016}. This relation points out that $F_8$ decreases with surface gravity. However, this technique can only be applied for stars with \logg $\geq$ 2.5 dex --since stars with lower \logg deviated from their relation --, and does not provide good granulation timescales for MS stars, since they have granulation timescales shorter than 30 minutes. \cite{Cranmer2014} used existing scaling relations to derive the star-integrated variability of a sample of \textit{Kepler} stars, and included an empirical correction factor for the magnetic suppression of convection in F-type stars. \cite{Pande2018} proposed a similar \textit{Flicker} method, but calculating instead the Fourier power spectrum to measure the granulation background to determine \logg \citep[see also][]{Ness2018}. Similarly, \cite{Bugnet2018} developed a slightly different tool called FliPer, which utilizes the average variability of a star measured in the power density spectrum at different frequency ranges. Additionally, \cite{Sulis2020} introduced a new indicator to track the granulation properties (the flicker index), which is defined as the power density spectrum (PSD) slope in the frequency regime dominated by this stellar signal. They found again strong correlation with the stellar parameters, such as \logg. \cite{VanKooten2021} improved the granulation flicker amplitude $F_8$ for a large sample of \textit{Kepler} stars, by incorporating the metallicity in the convective Mach number determination and using scaling relations derived from 3D models. 

\citetalias{Kallinger2014} analyzed the power spectra of red giants and MS stars observed by \textit{Kepler}, from which they extracted their granulation and global oscillation parameters. As a result, they found a tight relationship between the granulation parameters, \numax and \logg. They also note some disagreement with the predictions of \cite{Samadi2013}. Additionally, \cite{Kallinger2016} proposed a new method to determine a proxy for the characteristic timescale of the combined granulation and p-mode oscillation signal, as a tool to provide stellar surface gravities with accuracies of about 4\%.

The study and understanding of stellar granulation also aims to constrain the stellar noise present in light curves or radial velocity measurements\citep[see for example][]{Sulis2020b}. This because the granulation noise might impose problems to detect Earth-size or smaller exoplanets, and might affect the determination of exoplanet parameters such as the planetary radius or mass. Evidence of these was shown for example by \cite{Meunier2020}, who found that granulation and supergranulation contribute significantly to the uncertainties of the planetary mass of an Earth-mass planet, especially for high-mass stars, when using radial velocities. Additionally, \cite{Sulis2020} generated realistic exo-Earth transits on solar HMI observations, and demonstrated that the flicker noise could produce significant errors of up to 10\% on the planetary radius.

To improve the understanding of the relation between granulation and stellar parameters, we provide in this paper a better characterization of the stellar brightness fluctuations by using for the first time very long time series (i.e. more than 1000 convective turnover times), from 3D stellar atmosphere models distributed across the Hertzsprung-Russell (HR) diagram. Particularly, for the first time, we show that both the standard deviation and auto-correlation time scale in terms of a simple power law $\sim \nu_{\rm max}^\alpha$, and explore their dependencies with different metallicities. We found very good agreement when compared to observational Kepler data from \citetalias{Kallinger2014}, and a selected sample of stars with accurate stellar parameters.

The paper is structured as follows. In Section~\ref{sec:stagger} we described the STAGGER-code and how the 3D stellar atmosphere models are generated. The sample of 3D models is described in Section~\ref{sec:3Dsample} and a brief description of the code used to determine the radii of the 3D models is given in Section~\ref{sec:basta}. Section~\ref{sec:dataanalysis} is devoted to the determination of granulation properties using the 3D models, and the derivation of power-laws in terms of \numax. The discussion of the metallicity effect on the granulation properties is in Section~\ref{sec:Resultsmetallicities}, and a comparison of our solar metallicity models with \cite{Samadi2013} is in Section~\ref{sec:ComparisonSamadi}. In Section~\ref{sec:observations}, we present the comparisons with observational data, specifically with \citetalias{Kallinger2014} and selected stars from tne LEGACY sample \citep{Lund2017}. Finally, we discuss our results in Section~\ref{sec:Discussion} 
and conclude in Section~\ref{sec:conclusions}.

\section{The 3D hydro-simulations and their stellar parameters} 
\subsection{The STAGGER-code}
\label{sec:stagger}
 In this work, we used a standard version of the radiative-magnetohydrodynamic (R-MHD) STAGGER code \citep{Nordlund1995, Nordlund2009, Magic2013}, which generates 3D hydrodynamical model atmospheres that allow us to study stellar granulation across the HR diagram. The code solves the equations for the conservation of mass, momentum, and energy, as well as the radiative transfer equation under the assumption of local thermodynamic equilibrium (LTE). This equation is solved along a set of inclined rays specified by the user, which provides the heating and cooling rates needed for the energy equation. To account for the wavelength dependence of the radiative transfer we used the opacity binning method \citep{Nordlund1982, Skartlien2000, Ludwig2013, Collet2018} to generate bins with an averaged opacity strength over specific wavelength ranges. The line opacities are taken from \cite{Gustafsson2008}, while a sample of continuum absorption and scattering coefficients are adopted from \cite{Hayek2010}. Furthermore, a realistic treatment of microphysics is included in the code, which accounts for ionization and molecule formation. This is done by using an improved version of the Equation of State (EOS) by \cite{Mihalas1988} obtained by \cite{Trampedach2013} (see their Section 2.1), which contains the 17 most abundant elements found in the Sun, plus $\rm H_2$ and $\rm H^+_2$.   
 
Every 3D simulation created with the STAGGER-code is defined by three stellar parameters:  the entropy at the bottom that defines the effective temperature (\teff), surface gravity in logarithmic scale (\logg) and metallicity ([Fe/H]). \teff is determined by the entropy --defined by both energy per unit volume and density-- at the bottom boundary of every model, while the metallicity is defined by the abundance of chemical elements. The models cover the upper radiative atmosphere or photosphere, the superadiabatic region, and the quasi-adiabatic deeper convective layers, where we ensure a flat entropy profile at the bottom of the simulation domain. These layers are distributed in a specific 3D Cartesian geometry. This means that the models are centered around the photosphere, and thus, only the top layers of a star are modeled. 

\begin{figure}
	\resizebox{\hsize}{!}{\includegraphics{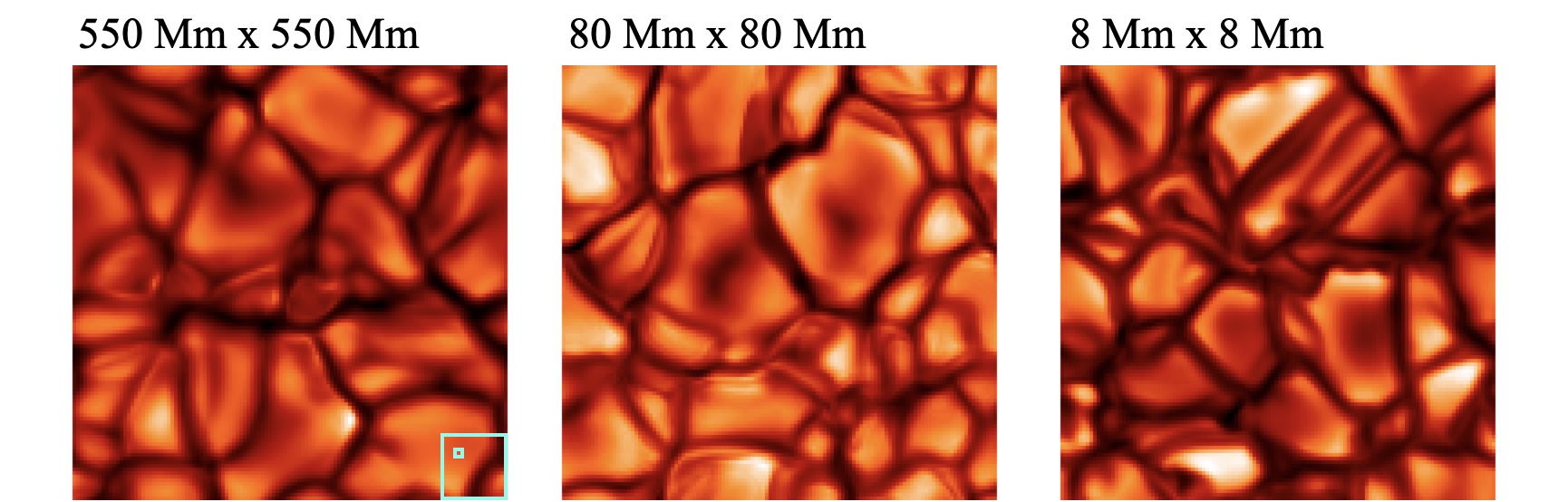}}
    \caption{Disk centered surface intensities at specific times of stellar atmosphere models at solar metallicity. \hspace{\textwidth}  \textit{Left:} a red giant \teff$ = 5000$ K, \logg$ = 2.50$. \textit{Middle:} a sub-giant \teff$ = 5500$ K, \logg$ = 3.50$. \textit{Right:} a solar-like star \teff$ = 5777$ K, \logg$ = 4.44$. The cyan scaled squares in the \textit{left} panel are representations of the surface intensities in the \textit{middle} and \textit{right} panels.}
    \label{fig:intensities}
\end{figure}

The simulation domain contains typically ten granular cells \citep{Magic2013} (see Fig.~\ref{fig:intensities} for some examples). The sizes are a few tens of surface pressure scale heights in all directions. Thus, the areas increase for models representing evolved stars or for the earlier type stars. Since the boxes are very small compared to the size of the stars, gravity is assumed to be constant in the entire box. The simulation domain is periodic horizontally, and with open boundaries vertically.

One of the quantities that can be retrieved from the 3D models is the 2D surface intensity $I(x,y,t)$, as function of the horizontal Cartesian coordinates  $x$ and $y$, and time $t$ of the snapshot. These intensities are used to compute the radiative bolometric flux defined as \citep[see, for example][]{Ludwig2006}

\begin{equation}
    \label{eq:bolflux}
    F(t) = 2\pi\sum_{i=1}^{N\rm_{bins}}\sum_{m=1}^{N_\mu}\sum_{k=1}^{N_\phi}w_{mk}(\cos{\theta})_m\avg{I^{(i)}_{mk}(x,y,t)},
\end{equation} 

where $\avg{I^{(i)}_{km}(x,y,t)}$ is the horizontal averaged intensity that corresponds to a given wavelength bin $(i)$ over the total number of bins $N\rm_{bin}$. The flux is the sum over various inclination cosines $(\cos{\theta})_m$ angles where $N_\mu$ indicates the number of rays, and $N_\phi$ the sum over the fraction of solid angle $w_{km}$ of the hemisphere at inclination $\theta_m$ and azimuthal angles over $k$. $w_{km}$ is defined by the Radau quadrature, multiplied by the constant weight over the azimuthal angle. For the 3D simulations used in this study we used $N_\mu = 2$ and $N_\phi =4$, and we justified this selection in Appendix~\ref{sec:numberrays}. 
 
 The 2D intensity $I_{km}^{(i)}(x,y,t)$ of a given bin $(i)$ is obtained by solving the (LTE) radiative transfer equation with the opacity binning method. Briefly, this method groups sets of wavelengths into a few bins based on the corresponding monochromatic continuous, line opacity, and their spectral interval \citep[see][]{Nordlund1982, Ludwig1992, Skartlien2000, Stein2003, Collet2018}, instead of solving the full monochromatic radiative transfer.

Fig.~\ref{fig:intensities} shows the bolometric intensity at a given time and for $\cos{\theta}=1$ for models representing a red giant star with \teff$= 5000 \ $K and \logg$= 2.50$, followed by a sub-giant star with \teff$= 5500$ K and \logg$= 3.50$, and a solar model. Each figure shows the granulation pattern, meaning bright hot granules surrounded by dark cooler intergranular lanes, which is a consequence of stellar convective up- and down-flows. The dimensions shown in the figure illustrate the fact that the granules increase in size as a star evolves, which is why the horizontal dimensions of our simulation boxes increase as well.

The 3D models also include box-modes \citep{Nordlund2001}, which are acoustic modes generated by both the stochastic excitation and the work done by the bottom boundary. These modes have, however, excessively large amplitudes compared to the ones observed in stars due to the shallow boxes and, consequently, they have very low inertia. Such large amplitude modes would affect the convective brightness simulations in an unrealistic way. We therefore decided to artificially damp them, so the time series are free of them, as explained in Appendix~\ref{sec:effects}. This is a good approximation, since in stars, granulation contributes about 2-4 times as much as the oscillations depending on the evolutionary stage of the star \citep[\citetalias{Kallinger2014};][]{Kallinger2016}.

Finally, we emphasize that the models belong to an improved version of the STAGGER-grid (Rodríguez Díaz, in prep.), which was originally published in \cite{Magic2013} with more than 200 models distributed across the HR diagram, with metallicities between +0.5 and -4.0. 

\subsection{Sample of 3D stellar atmosphere models}
\label{sec:3Dsample}
\begin{figure}
	\includegraphics[width=\columnwidth]{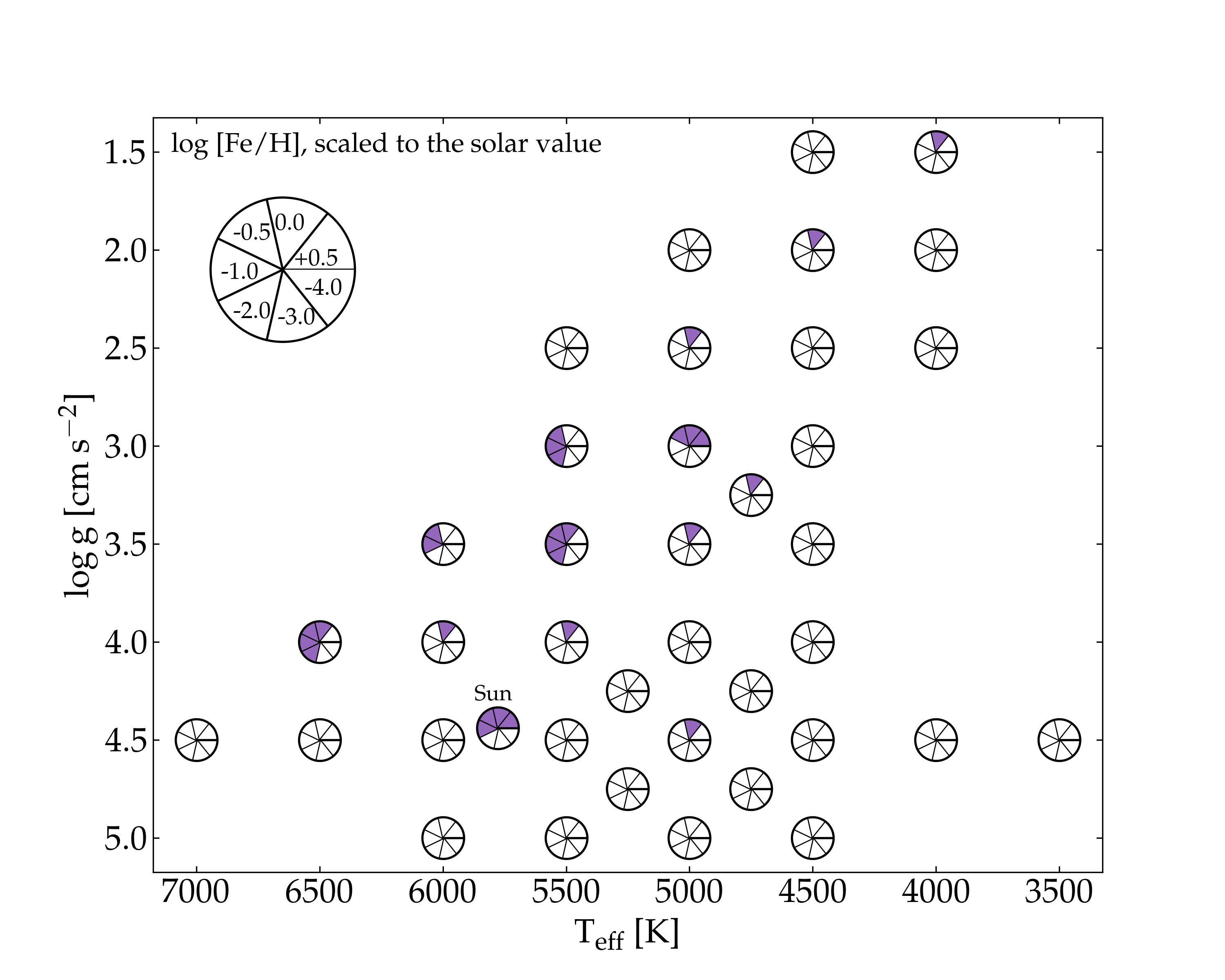}
    \caption{Kiel-diagram (\teff-\logg) showing an updated version of the STAGGER-grid. Every pie chart represents models with a given \teff and \logg, and metallicity decreases counterclockwise as illustrated in the figure. 
    The models highlighted in purple correspond to the models included in our sample, while models in white are the rest of the models in our grid.}
    \label{fig:gridp1}
\end{figure}

Our sample of 3D stellar atmosphere models includes K-dwarfs to giant stars with metallicities [Fe/H]$ = +0.5, 0.0, -0,5, -1.0, -2.0$,
where the abundances of $\alpha$-elements for models with [Fe/H] $\leq$ -1.0 include $\alpha$-enhancement of +0.4 dex. The models are distributed across the Kiel-diagram as illustrated in Fig.~\ref{fig:gridp1}. In Table~\ref{tab:sample} we summarized the stellar and physical parameters of our models, identified by their simulation names, which follow the nomenclature: \textit{txxgyymzz} or \textit{txxgyypzz}, where \textit{txx} indicates the first two digits of \teff, \textit{gyy} indicates \logg, and \textit{mzz} or \textit{pzz} indicate [Fe/H] $\leq$ 0.0 and [Fe/H] $>$ 0.0, respectively. 

Since the properties of convection change with stellar parameters, i.e. \logg and \teff, we used a characteristic quantity to select the models for this work, that is, according to their frequency $\nu_{\rm max}$. This quantity is defined as the ratio between sound speed to the pressure scale height, as $\nu_{\rm max}=c/H_p$, where \textit{c} scales roughly as (\teff/$\mu$)$^{1/2}$ with $\mu$ representing the mean molecular weight, and $H_p$ roughly as \teff/$g$. We emphasize here that this frequency is well-known in asteroseismology since it corresponds to the maximum of the power of acoustic modes. Even if our analysis does not contain modes, we keep this quantity as defined in asteroseismology since it follows nicely the evolution of the star $\nu_{\rm max} \sim g/\sqrt{T_{\rm eff}}$ \citep{Belkacem2011}. Low values of $\nu_{\rm max}$ correspond to evolved or hotter stars, while large values correspond to cool dwarfs. For the Sun its value is around $3090 \ \mu$Hz. Since we have models at different metallicities, we modified \numax according to the following dependencies: 

\begin{equation}
    \label{eq:molweight}
    \mu^{-1} = X + \frac{Y}{4} + \frac{Z}{\avg{A}},
\end{equation}
where $\mu$ is the mean molecular weight, \avg{A} corresponds to the mean atomic number, which is ${\approx}15.5$ for our models, and the mass fractions $X$, $Y$, and $Z$ were directly determined from the chemical abundances of the simulations, keeping in mind that the abundances of alpha-elements for [Fe/H]$\leq$-1.0 already accounted for $\alpha$-enhancement following 
\begin{equation}
    \label{eq:alphaenh}
    \text{[M/H]} = \text{[Fe/H]} + \text{corr([$\alpha$/Fe])},
\end{equation}

where \textit{corr} makes reference to a correction factor based on [$\alpha$/Fe] \citep{Salaris1993}.
Thus, the characteristic frequency \numax scales as \citep{Viani2017}

\begin{equation}
    \label{eq:numax_feh}
    \nu\rm_{max} \propto \left(\frac{\textit{g}}{\textit{g}_{\astrosun}}\right)\left({\frac{\textit{T}\rm_{{eff,\astrosun}} }{\textit{T}\rm_{eff}}}\right)^{1/2} \left(\frac{\mu}{\mu_{\astrosun}}\right)^{1/2},
\end{equation}

where $g_{\astrosun}, T\rm_{eff,\astrosun},$ and $\mu_{\astrosun}$ correspond to the solar values. We note that for our 3D models, we used the following reference values, based on the solar-like 3D model at solar metallicity: $g_{\astrosun} = 27542.29$ \ cm/s$^2$, $T\rm_{eff,\astrosun} =5759$ K, and $\mu_{\astrosun}=1.249$.
 We consider models of stars with 3 orders of magnitude range in \numax, as indicated in Table~\ref{tab:sample}. To calculate \numax for every model, we used $\avg{T\rm_{eff}}$, which is the mean \teff from the time series, rather than the target \teff corresponding to the \teff that we aimed to achieve for each model. Hereafter, unless specified, the \teff refers to $\avg{T\rm_{eff}}$. 

All 3D models in this work were generated with the following input physics: 

\begin{itemize}
    \item The time series covered at least 1000 convective turnover times, to have a reliable standard deviation independent of the length of the time series, opposite to what happens for too short time series of stochastic processes. 
    \item Opacity tables with 6 bins were used if available, otherwise 12 bins were used. We note that the influence of the number of bins is negligible as shown in Appendix~\ref{appendix:numberofbins}.
    \item The radiative transfer equation was solved along the vertical direction and 7 additional inclined angles. 
    \item We used a grid resolution of 120 points in every direction.
    \item The box-modes were damped.
\end{itemize}

Since we computed long time series, saved at relative high cadence, the relative low grid resolution was chosen to minimize both the CPU and storage resources, without affecting the derived granulation properties. This is justified in Appendix \ref{sec:effects}, where we studied the effects of our selection on the brightness fluctuations.

\subsection{The BAyesian STellar Algorithm (BASTA)}
\label{sec:basta}
Since the only stellar parameters that define the 3D models - \teff, \logg, and [Fe/H] - are indeed atmospheric parameters, it is not possible to describe a whole star with them. Therefore, several stars with different radii, masses, and ages may match a given set of these parameters. As we will discuss in Sec.~\ref{sec:dataanalysis}, to derive the \rms of the brightness fluctuations of an \textit{entire} star from a 3D model, we need to know its radius. In practice, to define the most \textit{probable} star that matches any of our models, we use the BAyesian STellar Algorithm  \citep[BASTA;][]{SilvaAguirre2015, SilvaAguirre2017, AguirreBoersenKoch2021}, a pipeline to accurately determine stellar properties using asteroseismology. A bayesian approach is implemented in BASTA to find the optimal solution from a pre-computed grid of evolutionary models, being in this study the grid of BaSTI isochrones \citep{Hidalgo2018} that included overshooting, diffusion, and mass loss. This is done by taking prior information such as a standard Salpeter initial mass function, and providing a set of observables, being in this case \teff, \logg, and [M/H]. [M/H] was calculated using Eq.~\ref{eq:alphaenh} with $\alpha$-enhancement [$\alpha$/Fe] = +0.4 dex when appropriated, as this is not the case for all metallicities. BASTA then takes all this information into account to compute the probability of every model in the grid matching the observations. As a result, BASTA provides a posterior probability distribution (PDF) for every parameter, and the final values for each stellar quantity are taken as a robust estimate of a typical star configuration that matches the 16 and 84 percentiles of the distribution. 

\begin{table*}
	\centering
	\caption{Stellar and physical parameters of the 3D stellar models used in this work. From left to right: model name, target effective temperature (target \teff) [K] corresponding to the \teff that we aimed to achieve for each model, surface gravity in logarithm scale ($\mathrm{\log{g}}$) [cm.s$^{-2}$], metalliciy [Fe/H], mean $\mathrm{\avg{T_{eff}}}$ [K] from the time series, stellar radius [$\mathrm{R_{\astrosun}}$] calculated with BASTA using isochrones, area of each local box model [Mm x Mm], brightness fluctuations (\rms) [ppm], and effective timescale (\taueff) [s].}
	\label{tab:sample}
	\begin{tabular}{lcccccccccr} 
		\hline
	   Model & Target $\mathrm{T_{eff}}$ [K] &  $\mathrm{\log{g} [cm.s^{-2}]}$ &  $\mathrm{\avg{T_{eff}}}$ [K]& [Fe/H] & Radius [$\mathrm{R_{\astrosun}}$] &  \numax [$\mathrm{\mu}$Hz] & (\textit{l}x\textit{l}) [Mm x Mm] & \rms [ppm] & \taueff [s] \\
		\hline
		t40g15m00   & 4000 & 1.50 & 3986 & 0.0  & 27.640$^{+0.821}_{-0.441}$  & 4.287    & 6600x6600 & 2139 & 150600 \\[0.5em]
		t45g20m00   & 4500 & 2.00 & 4522 & 0.0  & 21.600$^{+2.294}_{-5.137}$  & 12.727   & 2400x2400 & 1102 & 56730  \\[0.5em]
		t50g25m00   & 5000 & 2.50 & 4960 & 0.0  & 15.671$^{+0.612}_{-0.673}$  & 38.427   & 800x800   & 480  & 17160  \\[0.5em]
		t50g30p05   & 5000 & 3.00 & 4914 & 0.5  & 8.926$^{+0.862}_{-0.568}$   & 123.706  & 210x210   & 231  & 5328   \\[0.5em]
		t50g30m00   & 5000 & 3.00 & 4960 & 0.0  & 6.048$^{+0.356}_{-0.229}$   & 121.516  & 230x230   & 286  & 4852   \\[0.5em]
		t50g30m05   & 5000 & 3.00 & 4962 & -0.5 & 4.753$^{+0.191}_{-0.133}$   & 120.914  & 200x200   & 289  & 4800   \\[0.5em]
	    t55g30m05   & 5500 & 3.00 & 5508 & -0.5 & 7.734$^{+0.467}_{-0.402}$   & 114.719  & 250x250   & 215  & 4835   \\[0.5em]
	    t55g30m10   & 5500 & 3.00 & 5479 & -1.0 & 7.529$^{+0.148}_{-0.162}$   & 115.043  & 250x250   & 205  & 4858   \\[0.5em]
	    t55g30m20   & 5500 & 3.00 & 5545 & -2.0 & 6.733$^{+0.037}_{-0.087}$   & 114.226  & 220x220   & 154  & 4170   \\[0.5em]
		t47g32m00   & 4750 & 3.25 & 4727 & 0.0  & 3.613$^{+0.045}_{-0.046}$   & 221.351  & 88.8x88.8 & 220  & 2490   \\[0.5em]
		t50g35m00   & 5000 & 3.50 & 4958 & 0.0  & 2.886$^{+0.076}_{-0.077}$   & 384.345  & 65x65     & 153  & 1559   \\[0.5em]
		t55g35m00   & 5500 & 3.50 & 5516 & 0.0  & 3.834$^{+0.151}_{-0.230}$   & 364.387  & 80x80     & 145  & 1591   \\[0.5em]
		t55g35m05   & 5500 & 3.50 & 5545 & -0.5 & 3.560$^{+0.094}_{-0.093}$   & 361.888  & 64x64     & 118  & 1473   \\[0.5em]
		t55g35m10   & 5500 & 3.50 & 5430 & -1.0 & 3.251$^{+0.086}_{-0.387}$   & 365.435  & 70x70     & 112  & 1489   \\[0.5em]
		t55g35m20   & 5500 & 3.50 & 5485 & -2.0 & 2.533$^{+0.115}_{-0.033}$   & 363.186  & 60x60     & 97   & 1217   \\[0.5em]
		t60g35m05   & 6000 & 3.50 & 6004 & -0.5 & 3.614$^{+0.017}_{-0.091}$   & 347.751  & 81x81     & 140  & 1585   \\[0.5em]
		t60g35m10   & 6000 & 3.50 & 5912 & -1.0 & 3.419$^{+0.020}_{-0.072}$   & 350.221  & 80x80     & 127  & 1473   \\[0.5em]
		t60g40m00   & 6000 & 4.00 & 5962 & 0.0  & 1.792$^{+0.028}_{-0.028}$   & 1108.355 & 26x26     & 88   & 516    \\[0.5em]
		t65g40m00   & 6500 & 4.00 & 6413 & 0.0  & 1.985$^{+0.031}_{-0.035}$   & 1068.672 & 28x28     & 93   & 549    \\[0.5em]
		t65g40m05   & 6500 & 4.00 & 6485 & -0.5 & 1.761$^{+0.029}_{-0.026}$   & 1058.205 & 27x27     & 84   & 462    \\[0.5em]
		t65g40m10   & 6500 & 4.00 & 6447 & -1.0 & 1.758$^{+0.014}_{-0.027}$   & 1060.549 & 23x23     & 74   & 444    \\[0.5em]
		t65g40m20   & 6500 & 4.00 & 6440 & -2.0 & 1.651$^{+0.000}_{-0.036}$   & 1059.841 & 20x20     & 72   & 422    \\[0.5em]
		t55g40m00   & 5500 & 4.00 & 5462 & 0.0  & 1.646$^{+0.028}_{-0.027}$   & 1157.975 & 23x23     & 85   & 474    \\[0.5em]
		t5777g44p05 & 5777 & 4.44 & 5775 & 0.5  & 1.070$^{+0.003}_{-0.002}$   & 3142.903 & 8x8       & 54   & 194    \\[0.5em]
		t5777g44m00 & 5777 & 4.44 & 5759 & 0.0  & 0.994$^{+0.009}_{-0.013}$   & 3106.000 & 8x8       & 46   & 195    \\[0.5em]
		t5777g44m05 & 5777 & 4.44 & 5767 & -0.5 & 0.900$^{+0.010}_{-0.010}$   & 3090.649 & 8.3x8.3   & 40   & 188    \\[0.5em]
		t5777g44m10 & 5777 & 4.44 & 5743 & -1.0 & 0.787$^{+0.056}_{-0.047}$   & 3094.855 & 7.5x7.5   & 36   & 190    \\[0.5em]
		\hline
	\end{tabular}
\end{table*}

\section{Stellar convective noise from 3D stellar atmosphere models}
\label{sec:dataanalysis}

\begin{figure}
	\includegraphics[width=\columnwidth]{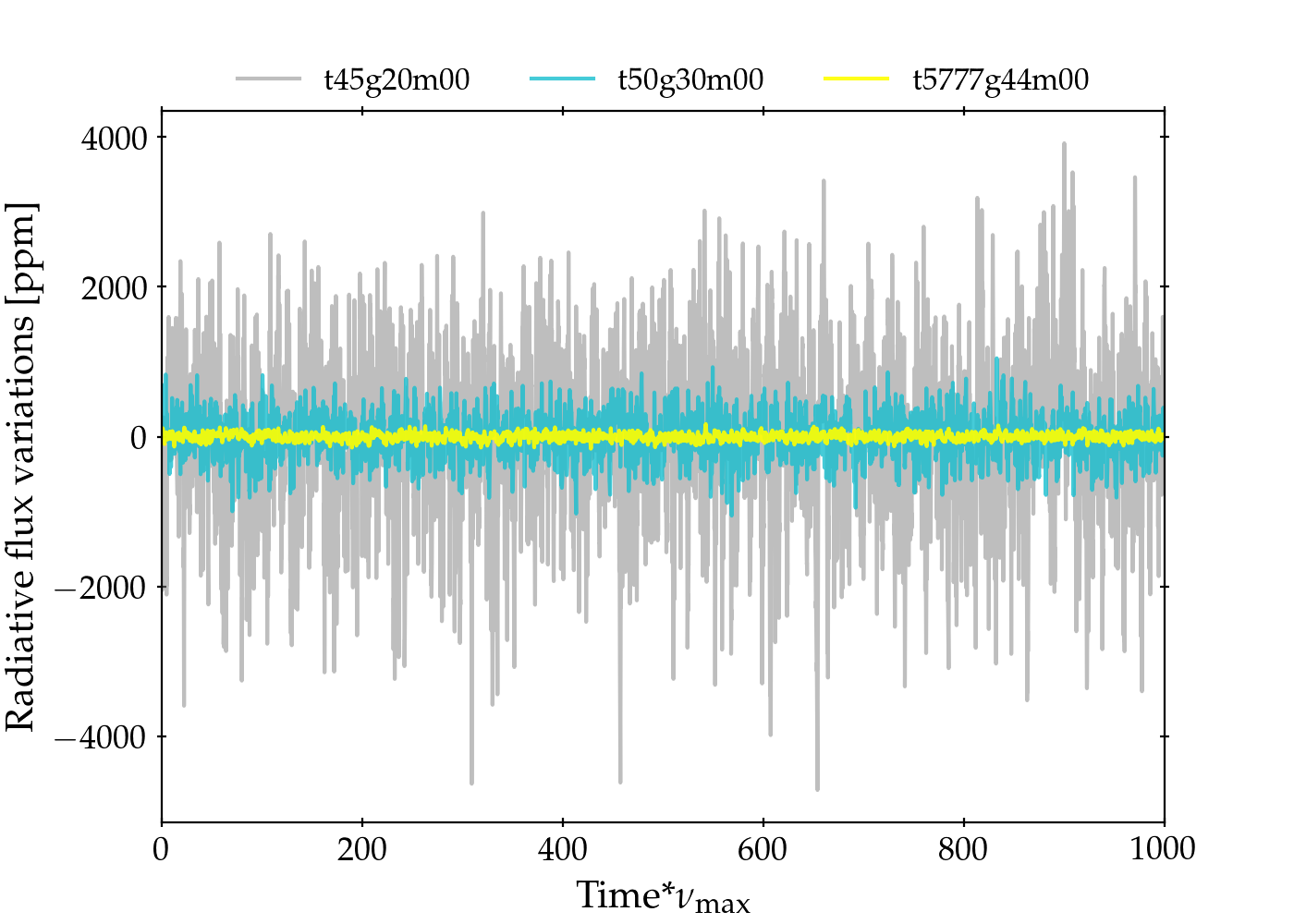} 
	\caption{Comparison of the radiative flux variations as a function of the number of convective turnover times, for three simulated stars at different evolutionary stages: giant (t45g20m00) with \numax$=12.727\mu$Hz, subgiant (t50g30m00) with \numax$=121.516\mu$Hz, and a solar simulation (t5777g44m00) with  with \numax$=3106 \mu$Hz.}
    \label{fig:fluxvar}
\end{figure}

Granulation produces fluctuations of the bolometric radiative flux as a function of time. Fig.~\ref{fig:fluxvar} shows models representing stars in the following stages: giant (t45g20m00), sub-giant (t50g30m00), and a solar simulation (t5777g44m00) in terms of the number of convective turnover times, defined here as the time from the time series scaled by the \numax associated with every model. From the figure it is clear, that the flux variations increase as a star evolves, and thus, increase with decreasing \numax.

This stochastic process can be described by two variables. First, its standard deviation $\mathrm{\sigma}$

\begin{equation}
    \label{eq:std}
        \sigma\rm_{box} =\sqrt{\avg{\mathcal{F}^2} - \avg{\cal F}^2},
\end{equation}
which describes the deviation of the bolometric radiative flux around its mean value. Second, the auto-correlation time (\taueff) of the signal,  which describes how long the fluctuations remain correlated in time. Our aim is to characterize these two quantities throughout the HR diagram, including their relation to fundamental stellar parameters and asteroseismic quantities. We note that the error bars associated to the power laws in this study correspond to standard errors of the fit, and thus, are 1-sigma error bars. This means that the errors only represent the deviation of our results with respect to the power laws.

\subsection{Power law of standard deviation}
\label{sec:ResultsSigma}
We first extracted the average value of the bolometric radiative flux $(\cal F)$ at each time step (integrated over the 6 or 12 opacity bins), which can then be used to determine \teff via the Stefan-Boltzmann law. 
The standard deviation ($\mathrm{\sigma_{box}}$) of the flux given by Eq.~\ref{eq:bolflux} is expressed in Eq.~\ref{eq:std}, which must be re-scaled by the number of granules \textit{N} visible on the disk, corresponding to the ratio of areas of visible disk to box surface, i.e.  $N =2\pi R_\star^2/l^2$, where $R_\star$ is the stellar radius, and $l$ is the length of one horizontal side of the 3D models \citep[see][]{Trampedach1998, Ludwig2006}:

\begin{equation}
    \label{eq:ratio}
        \sigma = \sigma_{\text{box}}\sqrt{N^{-1}}. 
\end{equation}

This relation assumes that the stochastic process is ergodic and that all patches are statistically uncorrelated.

To determine \rms, we first randomized the $\cal F$ values of each time series by shuffling the order of the data points, to emulate the kind of signal a telescope will receive, that is a signal with contributions from the whole stellar disk. Subsequently, we calculated \textit{\rms} at cumulative intervals of five hours, to determine the value towards which \rms converges.

\begin{figure}
	\includegraphics[scale=0.45]{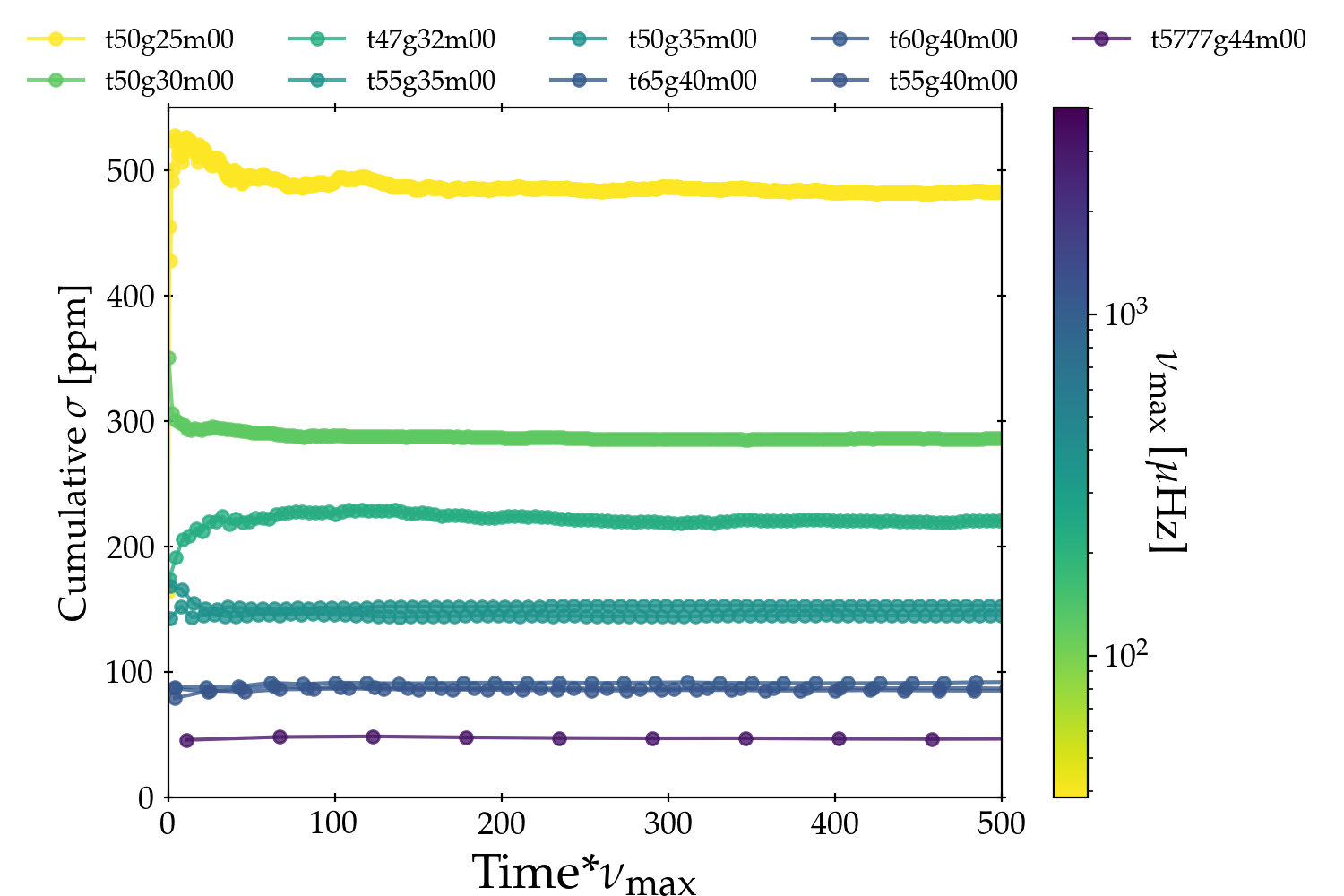}
    \caption{Cumulative standard deviation \rms of the relative brightness fluctuations for the 3D stellar atmosphere models at solar metallicity, as a function of the number of convective turnover times. Each point in the plot represents the cumulative value of \rms from 0 to the end of the time series, at intervals of five hours.}
    \label{fig:RMSsample}
\end{figure}

Fig.~\ref{fig:RMSsample} shows \rms for some 3D stellar atmosphere models at solar metallicity, in terms of the number of convective turnover times. It can be seen that the brightness fluctuations increase as the star evolves towards the turn-off point and Red Giant Branch. For all targets, we found that at least $300$ turnover convective times are needed to obtain a reliable \rms value. This means that RHD 3D simulations need to be long to perform reliable statistical studies of this noise source. 

\begin{figure*}
  \centering
  \begin{subfigure}[tb]{0.49\textwidth}
    \includegraphics[width=\columnwidth]{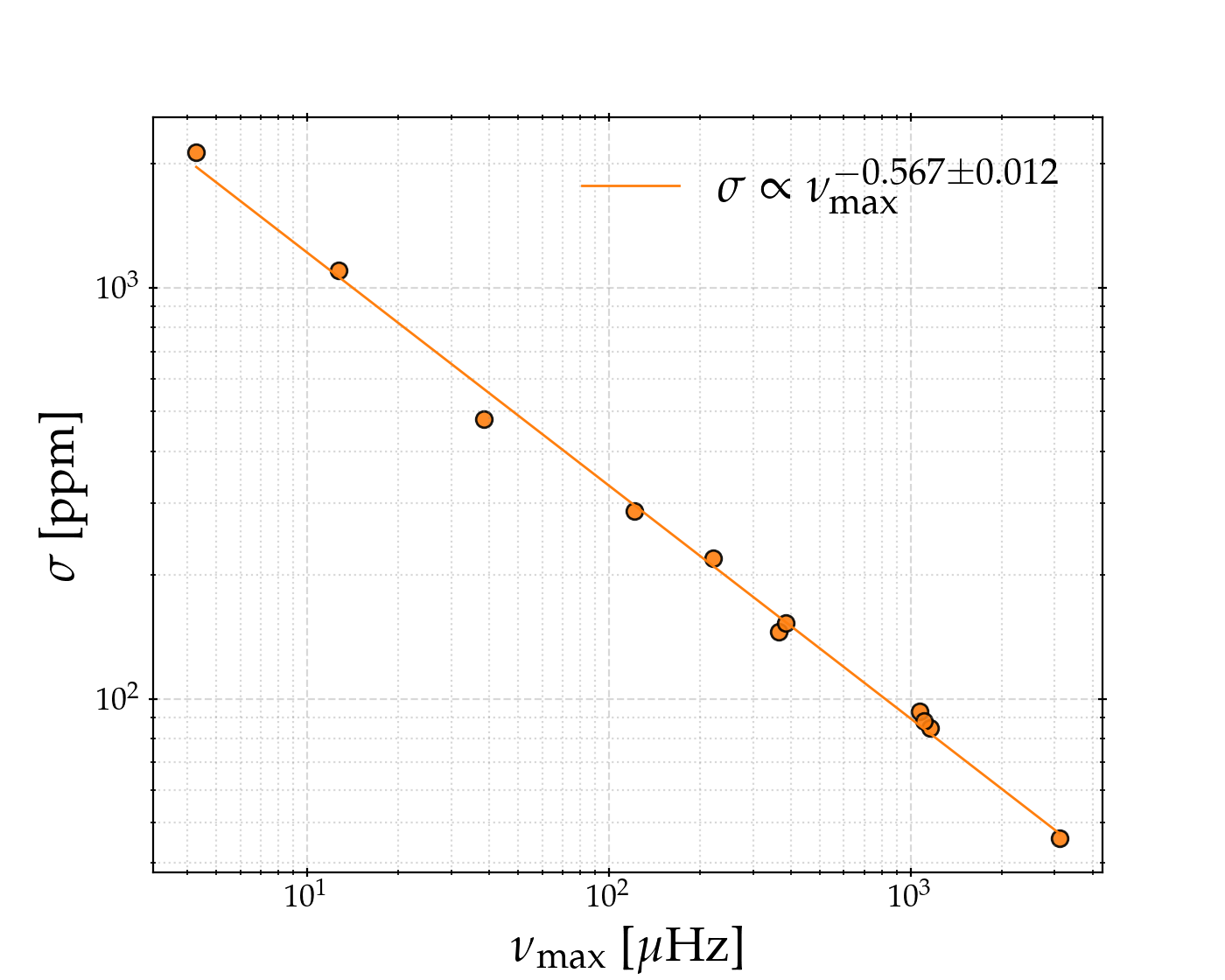}
  \end{subfigure}
  \hfill
  \begin{subfigure}[tb]{0.49\textwidth}
    \includegraphics[width=\columnwidth]{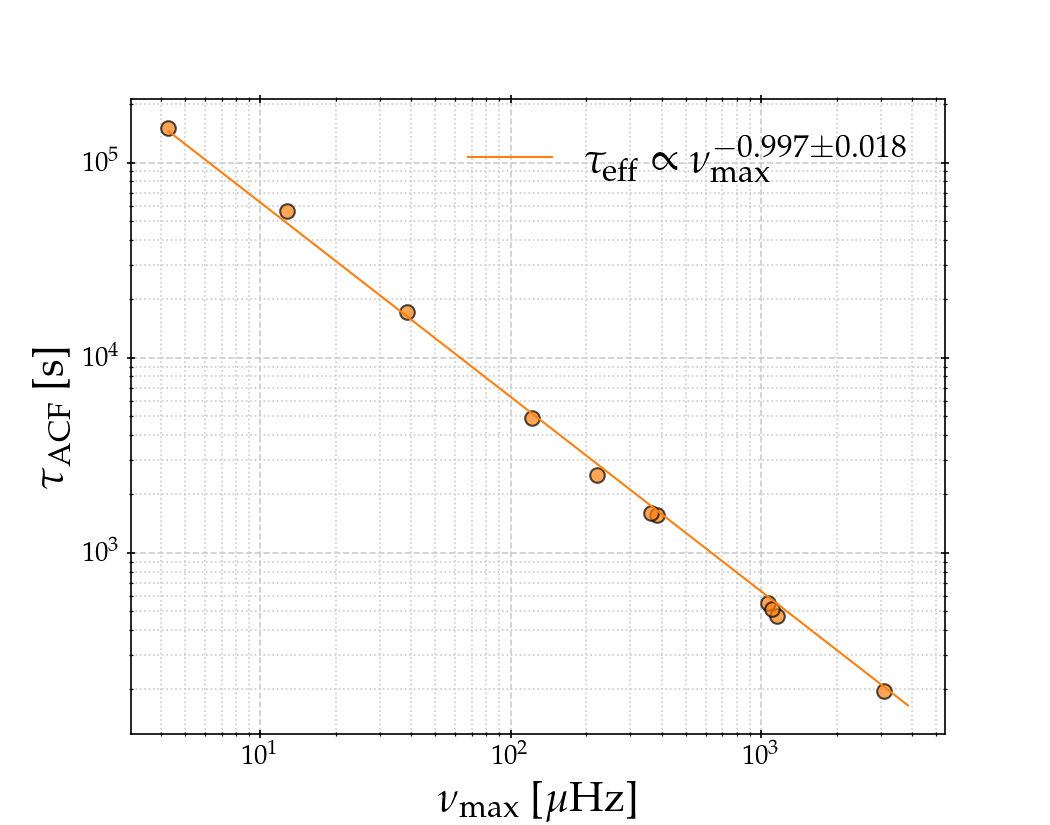}
  \end{subfigure}
  \caption{Standard deviation \rms (left) and auto-correlation time \taueff (right)  for our sample of 3D models at solar metallicity, as functions of \numax.}
  \label{fig:SampleResults}
\end{figure*}

In order to analyze the relationship between the \rms and stellar parameters, Fig.~\ref{fig:SampleResults} shows the \rms of our models at solar metallicity in terms of \numax. The figure reproduces the fact that \rms increases as \numax decreases, as we have previously pointed out, and as some previous studies have shown \citep[i.e][\citetalias{Kallinger2014}]{Mathur2011, Samadi2013}. However, our results show for the first time a very tight correlation expressed in terms of a single exponent of \numax, i.e. $\sim \nu_{\rm max}^\alpha$, without the need to introduce an additional quantity (e.g. Mach number). We will expand more on this in Sec.~\ref{sec:ComparisonSamadi}.  The power-law fit to our data indicates that

\begin{equation}
    \sigma \propto \nu\rm_{max}^{-0.567\pm 0.012}.
\end{equation}
The slight deviations of individual models from the power-law are due to the determination of the radii, since their determination with BASTA is slightly less precise when using only the atmospheric parameters \teff, \logg, and [Fe/H] \citep[see Fig. 8 in][]{AguirreBoersenKoch2021}.

\subsection{Characteristic time scale \texorpdfstring{$\tau\rm_{ACF}$}{tauACF}}
\label{sec:ResultsTime}

We define the characteristic time of our time series as the auto-correlation time \taueff, i.e. the time during which the fluctuations remain correlated in time. This is closely related to the convective turnover time. The auto-correlation function (ACF) of our time series is determined as the inverse Fourier transform of the PSD of the time series. \taueff is defined to be the e-folding time of the ACF \citep[see e.g.][]{Mathur2011} In Fig.~\ref{fig:SampleResults} we show the tight relation between \taueff and \numax, where it is clear that the more evolved the star is, the longer the timescale is. The power law shows that 

\begin{equation}
    \label{eq:powerlaw-time}
   \tau\rm_{ACF} \propto \nu\rm_{max}^{-0.997 \pm 0.018}.
\end{equation}

These 3D numerical results confirm the $\sim 1/\nu_{\rm max}$ proposed by \cite{Kjeldsen2011}, based on semi-empirical relation. It is indeed not surprising to find such dependence if we assume that the convective speed scales like the sound speed ($c$) and the characteristic size scales like the pressure scale ($H_p$), therefore the characteristic time must scale like $\sim H_p/c \sim 1/\nu_{\rm max}$. 

\subsection{Effects of metallicity}
\label{sec:Resultsmetallicities}

The granulation properties change with metallicity which have been found by numerical simulations \citep[e.g.][]{Nordlund1990, Magic2013, Ludwig2016} or from observations  \citep[e.g. for evolved stars][]{Corsaro2017, Yu2018}. Until now, the effects of metallicity on brightness fluctuations has never been investigated using 3D simulations with more than two different metallicities, which is what we explore in this section. We focus on  how these brightness fluctuations change with the stellar parameters  and metallicity.

\begin{figure*}
  \centering
  \begin{subfigure}[tb]{0.49\textwidth}
    \includegraphics[width=\columnwidth]{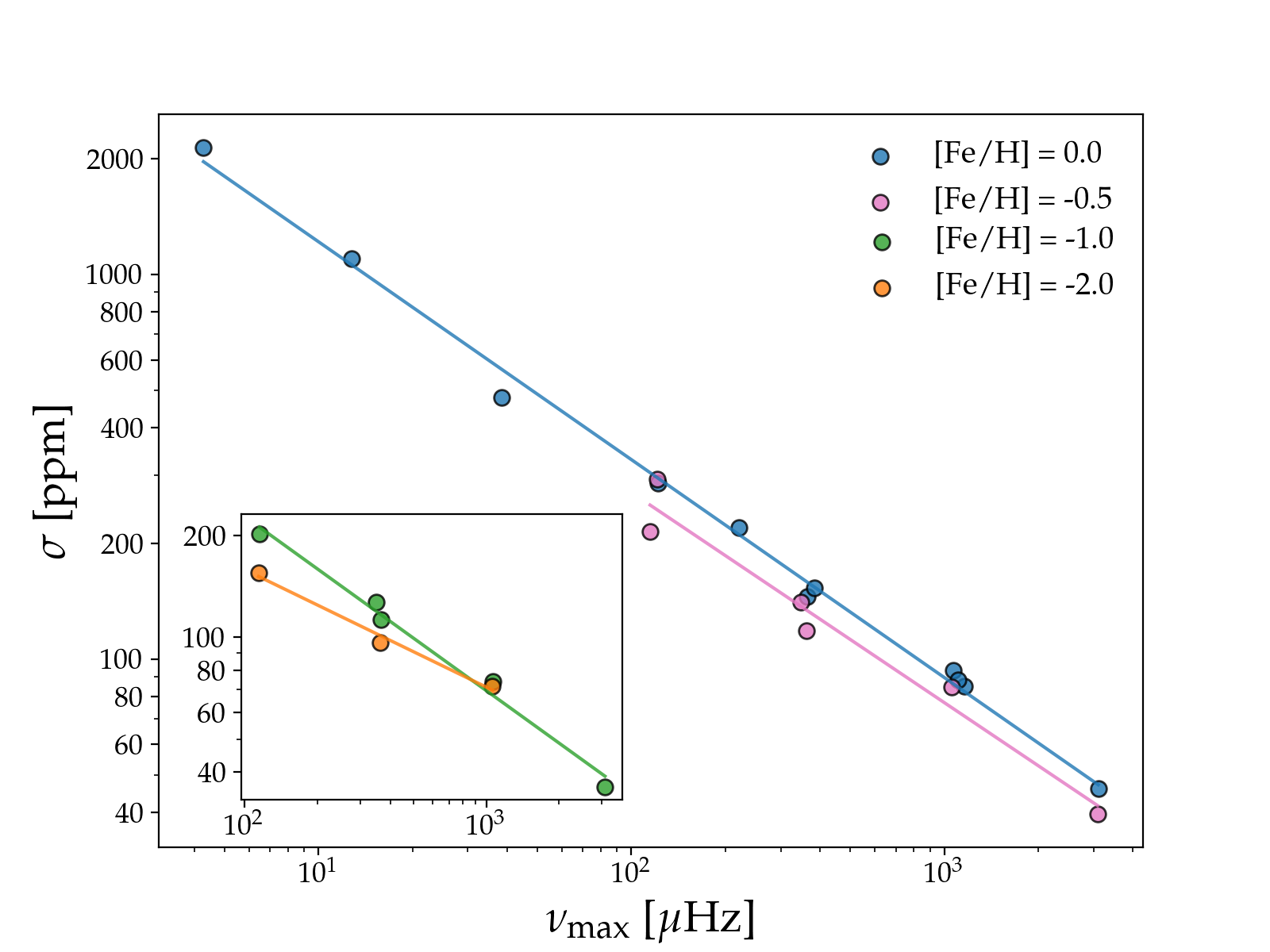}
  \end{subfigure}
  \hfill
  \begin{subfigure}[tb]{0.49\textwidth}
    \includegraphics[width=\columnwidth]{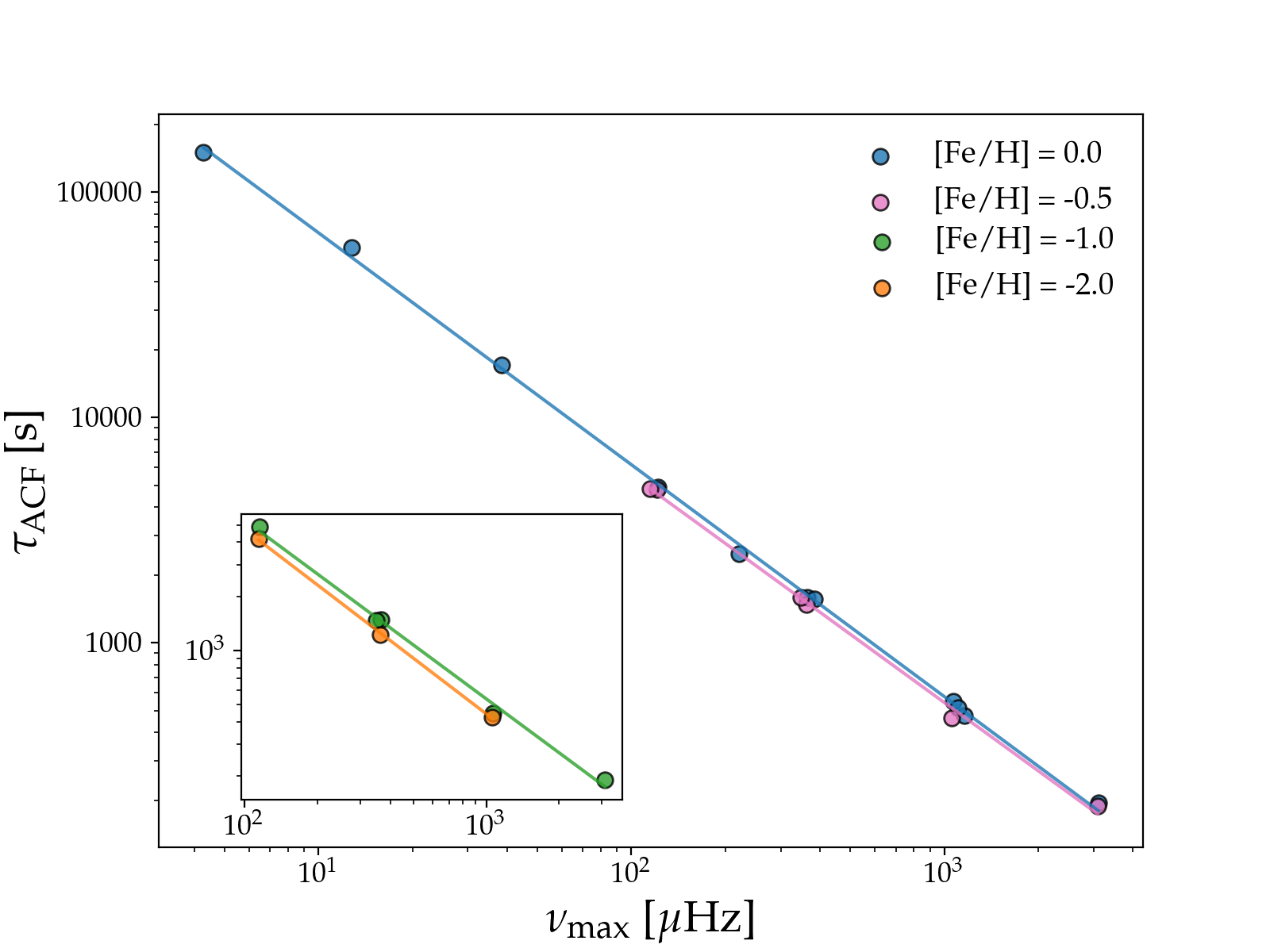}
  \end{subfigure}
  \caption{Standard deviations \rms (left) and auto-correlation time  \taueff (right) as a function of metallicity. We clearly show a power law dependence between \rms and \numax in terms of [Fe/H] (see text), where the coefficients decrease with metallicity.}
  \label{fig:CompMetallicities}
\end{figure*}

In Fig.~\ref{fig:CompMetallicities} we show the relation between \rms and \taueff with \numax for every individual metallicity, i.e. 0.0, -0.5, -1.0, and -2.0. We excluded [Fe/H] = +0.5 since our sample only contained two models at that metallicity. In Table~\ref{tab:valuesfitFeH}
we summarized the power law relations between the granulation properties and \numax. For the first time, 3D models are used to show that metallicity does have a significant impact on \rms, as also found by \cite{Corsaro2017, Yu2018} using observations of evolved stars. We emphasize that they only considered evolved stars in their analysis and were restricted to a narrower range of metallicities. Specifically, we clearly show that \rms decreases with metallicity, which is easily explained since the sizes of the granules also decrease with metallicity \citep{Collet2007,Magic2013}. Therefore, $\alpha$ in \rms $\propto \nu\rm_{max}^\alpha$ decreases with metallicity. We want to mention that the model at highest \numax at [Fe/H] = -2.0 has a higher \rms than expected, perhaps due to to the horizontal dimensions of the model, which is why the trend looks different at that metallicity. 

We fitted also a function of the form

\begin{equation}
    \label{eq:sigmaFeH}
    \ln{\left(\frac{\sigma}{\sigma_{\odot, 3D}}\right)} = s\ln{\left(\frac{\nu\rm_{max}}{\nu\rm_{max, 3D}}\right)} + u\text{[Fe/H]},
\end{equation}

similar to the fit used by \cite{Corsaro2017}, where s and u are the constant to fit. We did not include the mass dependency as they did, since they used scaling relations to determine the mass for their targets, whereas the mass for the 3D models is determined with BASTA. We disregard the propagation of errors due to the determination of the stellar mass, therefore we do not include it in our results for simplicity. But using this equation to fit all the 3D models of our sample, we found that $s = -0.556 \pm 0.010$, and $u = 0.211 \pm 0.033$.

The relation derived by \citet{Corsaro2017} ($u = 0.89 \pm 0.08$) was obtained for a restricted part of the power spectrum -at low frequencies- that they defined as meso-granulation. Despite the fact that we have very long simulated time series, they are still much shorter than the time series from \textit{Kepler}. Therefore, in most of our simulations it was hard to see the different components of granulation in the power spectrum, or difficult to obtain a stable fit with Harvey laws. Instead, we extracted \rms by computing the standard deviation in the time domain and not in the Fourier space. For this reason, we emphasize that our estimation of $\sigma$ might slightly differ from those presented in \citet{Corsaro2017}. 

This explains why our $u$ value is approximately 4 times lower than their value. We want to point out though, that they did not include the metallicity effect from the bolometric correction \citep[see][]{Lund2019}, which might have a small impact on their derived $u$ parameter. In conclusion, we provide power-law relations between \rms and \numax for four different metallcities, and a general relation that has metallicity as one of its parameters, so that the reader can decide which one to use.

\begin{table}
	\centering
	\caption{Values for the relation between \rms and \taueff with \numax for every [Fe/H], as shown in Fig.~\ref{fig:CompMetallicities}.}
	\label{tab:valuesfitFeH}
	\begin{tabular}{lcc} 
		\hline
	    [Fe/H] &  \rms $\propto \nu\rm_{max}^\alpha$ &  \taueff $\propto \nu\rm_{max}^\beta$  \\
		\hline
        0.0  & -0.567 $\pm$ 0.012 & -1.029 $\pm$ 0.009\\
        -0.5 & -0.546 $\pm$ 0.053 & -1.009 $\pm$ 0.025\\
        -1.0 & -0.516 $\pm$ 0.034 & -0.997 $\pm$ 0.039\\
        -2.0 & -0.348 $\pm$ 0.038 & -1.029 $\pm$ 0.022\\

		\hline
	\end{tabular}
\end{table}

Additionally, we explore in the right panel of Fig.~\ref{fig:CompMetallicities} the relation between \taueff and \numax for four different metallicities as well. Overall, \taueff decreases slowly with metallicity, as also seen in the $\beta$ values in Table~\ref{tab:valuesfitFeH} for the relation \taueff $\propto \nu\rm_{max}^\beta$, where it is clear that the timescales show a weak dependence on metallicity.

\subsection[comparison with]{Comparison with Samadi et al. (2013)} 
\label{sec:ComparisonSamadi}
\cite{Samadi2013} studied two granulation properties (\rms and \taueff) of 3D models at solar metallicity using the CIFIST grid \citep{Ludwig2009b}. They adopted a different method to analyze the models compared to the method used in this work, where they fitted the power spectrum of the times series using Harvey laws to extract their parameters. 

\begin{figure*}
  \centering
  \begin{subfigure}[tb]{0.49\textwidth}
    \includegraphics[width=\columnwidth]{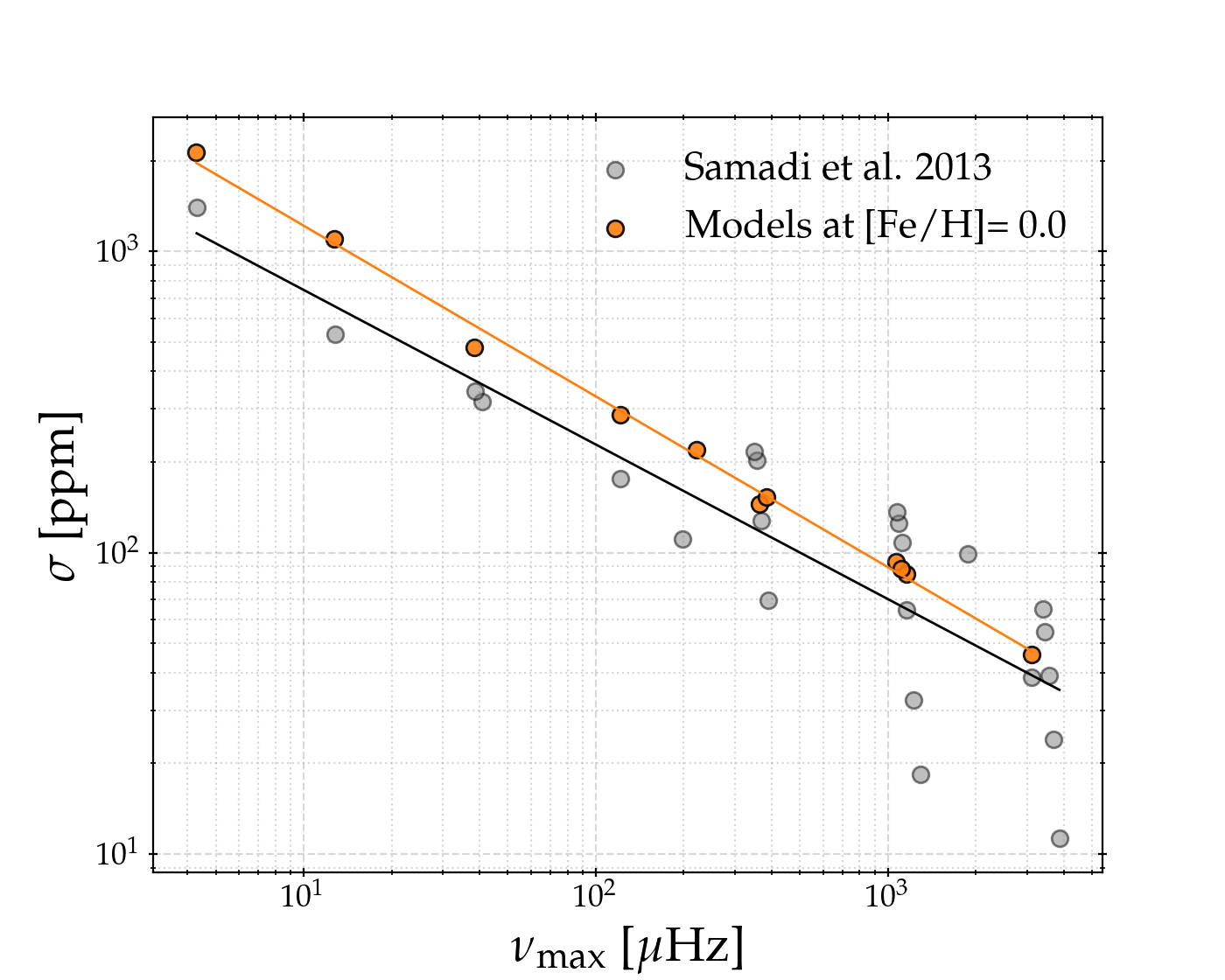}
  \end{subfigure}
  \hfill
  \begin{subfigure}[tb]{0.49\textwidth}
    \includegraphics[width=\columnwidth]{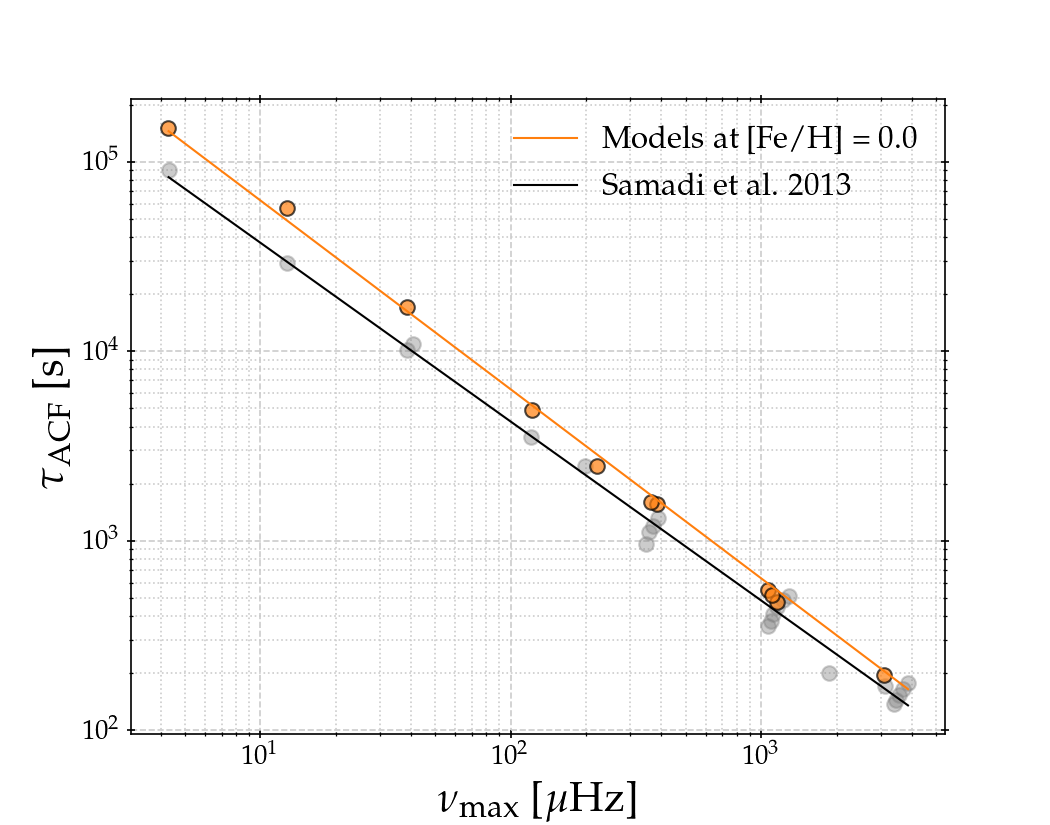}
  \end{subfigure}
  \caption{Comparison of standard deviations \rms (Left)  and auto-correlation times \taueff (Right), as functions of \numax, between our sample and the one presented by \citet{Samadi2013}. The orange and black power law fits correspond to the fits of our data and \citet{Samadi2013} data, respectively, whose information is listed in Table~\ref{tab:relationsSamadi}.}
  \label{fig:CompSamadi}
\end{figure*}

\begin{table}
	\centering
	\caption{\textit{$\alpha$} values from the relation between \rms and \numax and between  \taueff and \numax from Fig.~\ref{fig:CompSamadi}, for the 3D models of this study and the 3D models of \citet{Samadi2013}}
	\label{tab:relationsSamadi}
	\begin{tabular}{lcc} 
		\hline
	    Relation & \textit{$\alpha$} - This study & \textit{$\alpha$} - \citep{Samadi2013}\\
		\hline\hline
	    \rms $\propto$ $\mathrm{\nu_{max}^\alpha}$& -0.567 $\pm$ 0.012 & -0.513 $\pm$ 0.062\\
	    \taueff $\propto$ $\mathrm{\nu_{max}^\alpha}$& -0.997 $\pm$ 0.018 & -0.944 $\pm$ 0.019\\
	    \hline
	\end{tabular}
\end{table}

Fig.~\ref{fig:CompSamadi} shows \rms as a function of \numax for the 3D models for the two analyses. In each case, we  determined the relation \rms $\propto$ $\mathrm{\nu_{max}^\alpha}$ for both samples, which are summarized in Table~\ref{tab:relationsSamadi}. Despite the fact that the power $\alpha$ in the two studies are within the 1-sigma agreement (see Table~\ref{tab:relationsSamadi}), the \rms values determined by \cite{Samadi2013} show a strong scatter that we do not observe in our sample, allowing us to associate a unique \rms value for a given \numax, something that was not possible in \cite{Samadi2013}. In order to derive a power-law like behavior they introduced a different reference quantity $\sim z_3^\alpha$ which accounts for the Mach number, in order to decrease the scatter of the \rms values. Our results show that there is no need to introduce such dependence, and \rms can be derived with only knowing the stellar parameters. 

Given that both the STAGGER and CO$^5$BOLD codes have similar physics and structures, we attribute the different behavior to two main reasons: their short time series and their radius determination. Since their time series were only about 9 hours long for solar-like stars (Ludwig H.-G. private comm.), this might introduce an incorrect \rms by about 2-6\%, based on the early fluctuations associated with \rms in Fig.~\ref{fig:relativeRMSsample}. Nevertheless, these fluctuations do not account for all the differences seen in Fig.~\ref{fig:CompSamadi}. Regarding the stellar radii, they did not provide enough information about how the values were determined. Nevertheless, we know that those values depend on the atmospheric parameters of the 3D models, and the method and underlying physical assumptions used to calculate the radii. If the proper conditions are not chosen, the code might provide an incorrect stellar radius. By comparing the radii values from \cite{Samadi2013} and the ones determined in this work with BASTA for the models with similar or equal \teff, \logg, and [Fe/H], we found that the values reported by \cite{Samadi2013} are overestimated, especially for models with \logg $\leq$ 3.0. For example, for the t40g15m00 models, the radius difference is of $12.18R_\odot$, while for the t50g25m00 models the difference is $2.579R_\odot$. Another possible reason that might contribute to the observed scatter in \citet{Samadi2013} is, that the box modes were not damped (Ludwig, Private comm) in their study, opposite to our approach.

Regarding the timescales \taueff, we show in Fig.~\ref{fig:CompSamadi} \taueff of our solar metallicity models with the values reported by \cite{Samadi2013}, as a function of \numax. The figure shows that we recover the same trend. In fact, our relation \taueff(\numax) shown in Eq.~\ref{eq:powerlaw-time} is in better agreement with the relation derived by \cite{Kjeldsen2011}, i.e. \taueff $\propto$ $\mathrm{\nu_{max}^{-1}}$. Even though both power laws are very similar, it is worth mentioning that \cite{Samadi2013} data for \numax higher than 120 $\mu$Hz present a relatively significant scatter of up to $\sim 500$ s, contrary to the data presented in this study. Therefore, we can again provide a unique value of \taueff for a given set of stellar parameters.

\section{Comparison with \textit{Kepler} stars}
\label{sec:observations}
We aim to compare our theoretical predictive power laws with a sample of stars observed by the \textit{Kepler} space mission, which observed more than 145,000 stars for more than four year. The observations were either carried out in long cadence (LC) of ${\sim}30$ min or short cadence (SC) of ${\sim}1$ min. We refer to \cite{Borucki2010} for more detailed information about the Kepler mission.  
\\\\
In this study, we used two different samples of \textit{Kepler} stars, which will be introduced below. The first sample corresponds to the stars analyzed in \citetalias{Kallinger2014} (see their Sec. 2), where they focused on LC data of 1289 red-giant stars that had previously been used in other studies (see references therein), and SC data of 75 sub-giant and solar-type stars, originally presented in \cite{Chaplin2011}. The stellar parameters of these stars were obtained from scaling relations. The second sample corresponds to the LEGACY stars \citep{Lund2017}, which are MS stars with frequency-power spectra optimized for asteroseismology. The selected stars together with their stellar parameters are shown in Table.~\ref{tab:keplersample}. We emphasize that the stellar parameters obtained from the fit of the individual mode frequencies are more accurate than those obtained from simple scaling relations.
For the LEGACY sample, we used the data in \citet{Lund2017}. We point out, that three stars of the LEGACY sample are in common with the sample of \citetalias{Kallinger2014}, specifically KIC 6106415, KIC 5184732, and KIC 12317678. 

\begin{table*}
	\centering
	\caption{Stellar properties and statistical uncertainties of selected stars from the LEGACY sample \citep{Lund2017}. \rms values were determined by fitting their power spectrum.  An example of this fitting is shown in Appendix~\ref{sec:psdfitting}}.
	\label{tab:keplersample}
	\begin{tabular}{lcccccc}
		\hline
	   KIC  & \teff [K] & \logg [cm.s$^{-2}]$ & [Fe/H] & \rms [ppm] & \taueff [s] & \numax [$\mu$Hz] \\
		\hline \hline
		  \multicolumn{7}{c|}{ Sample of LEGACY stars}  \\
            \hline 
            3656476	 & 5668 $\pm$ 77     & 4.225$_{-0.008}^{+0.010}$  &  0.25 $\pm$ 0.10    & 59.162$_{-0.310}^{+0.298}$ & 308.394$_{-4.803}^{+4.662}$   & 1925 \\[0.5em]
            5184732	 & 5846 $\pm$ 77     & 4.255$_{-0.010}^{+0.008}$  &  0.36 $\pm$ 0.10    & 55.078$_{-0.143}^{+0.149}$ & 297.757$_{-3.484}^{+3.496}$   & 2089 \\[0.5em] 
            6106415  & 6037 $\pm$ 77     & 4.295$_{-0.009}^{+0.009}$  & -0.04 $\pm$ 0.10    & 45.647$_{-0.138}^{+0.145}$ & 262.891$_{-3.709}^{+3.499}$   & 2249 \\[0.5em]  
            6679371	 & 6479 $\pm$ 77     & 3.934$_{-0.007}^{+0.008}$  &  0.01 $\pm$ 0.10    & 67.988$_{-0.494}^{+0.533}$ & 510.592$_{-19.017}^{+16.871}$ & 942  \\[0.5em]
            7106245	 & 6068 $\pm$ 102(3) & 4.310$_{-0.010}^{+0.008}$  & -0.99 $\pm$ 0.19(3) & 44.448$_{-1.578}^{+1.487}$ & 254.292$_{-25.214}^{+25.088}$ & 2398 \\[0.5em]
            7296438	 & 5775 $\pm$ 77     & 4.201$_{-0.009}^{+0.010}$  & 0.19 $\pm$ 0.10     & 58.894$_{-0.505}^{+0.524}$ & 313.926$_{-8.071}^{+9.521}$   & 1848 \\[0.5em]
            7871531	 & 5501 $\pm$ 77     & 4.478$_{-0.007}^{+0.005}$  & -0.26 $\pm$ 0.10    & 32.139$_{-0.294}^{+0.257}$ & 244.392$_{-5.699}^{+6.820}$   & 3456 \\[0.5em]
            7970740	 & 5309 $\pm$ 77     & 4.539$_{-0.004}^{+0.005}$  & -0.54 $\pm$ 0.10    & 30.866$_{-0.119}^{+0.123}$ & 232.506$_{-2.423}^{+2.666}$   & 4197 \\[0.5em]
            8006161	 & 5488 $\pm$ 77     & 4.494$_{-0.007}^{+0.007}$  & 0.34 $\pm$ 0.10     & 42.786$_{-0.378}^{+0.862}$ & 186.484$_{-4.209}^{+15.512}$  & 3575 \\[0.5em]
            9025370	 & 5270 $\pm$ 180(2) & 4.423$_{-0.007}^{+0.004}$  & -0.12 $\pm$ 0.18(2) & 30.047$_{-0.282}^{+0.266}$ & 227.339$_{-4.565}^{+5.452}$   & 2989 \\[0.5em]
            9098294	 & 5852 $\pm$ 77     & 4.308$_{-0.005}^{+0.007}$  & -0.18 $\pm$ 0.10    & 46.229$_{-0.473}^{+0.451}$ & 253.942$_{-7.461}^{+7.846}$   & 2315 \\[0.5em]
            12069127 & 6276 $\pm$ 77     & 3.912$_{-0.005}^{+0.004}$  & 0.08 $\pm$ 0.10     & 63.718$_{-1.613}^{+2.292}$ & 489.342$_{-67.995}^{+49.358}$ & 885  \\[0.5em]
            12258514 & 5964 $\pm$ 77     & 4.126$_{-0.004}^{+0.003}$  & -0.00 $\pm$ 0.10    & 56.478$_{-0.154}^{+0.153}$ & 374.428$_{-5.737}^{+5.441}$   & 1513 \\[0.5em]
            12317678 & 6580 $\pm$ 77     & 4.048$_{-0.008}^{+0.009}$  & -0.28 $\pm$ 0.10    & 57.130$_{-0.401}^{+0.461}$ & 353.875$_{-14.398}^{+10.962}$ & 1212 \\[0.5em]
		\hline
	\end{tabular}
\end{table*}

\subsection{Determination of granulation properties of our selected stars (LEGACY)} 
\label{sec:RMSKepler}
To determine the \rms values for the selected sample of LEGACY stars, we decided to fit the background model of the power spectra, instead of using the power-law relations derived in the previous section. This with the aim of verifying that different techniques can yield similar results, and thus, use observations to validate the power-laws derived using the 3D stellar atmosphere models. 

The power spectrum was calculated from a weighted least-square sine-wave fitting to the time series \citep[see][for more details]{kasoc_filter, Lund2017}.
We fit for each star a background model of the power spectrum of the form \citep[][\citetalias{Kallinger2014}]{Harvey1985}
\begin{equation}\label{eq:bgfunc}
    P(\nu) = P_n + \eta^2(\nu) \left[ \sum_{i=1}^3 \frac{\xi_i \sigma_i^2 \tau_i}{1 + (2\pi\nu\tau_i)^{a_i}}  + G(\nu)\right].
\end{equation}

Here $\eta(\nu)$ describes the apodization of the signal amplitude (square root of power) at frequency $\nu$ from the finite sampling, and in effect binning, of the temporal signal \citep{Chaplin2011b}; $\tau_i$ gives the characteristic time scale of the $i$th background component; $\sigma_i$ gives the corresponding \rms variation of the component in the time domain; $P_n$ denotes a constant shot-noise level; $a_i$ gives the slope of the decay, and thus the amount of memory in the background phenomenon (a value of $2$ corresponds to a standard Harvey model);  $\xi_i$ is a normalisation constant such that the integral of the background equals $\sigma_i^2$ in accordance with Parseval's theorem \citep[see, e.g.,][]{Michel2009}. To determine the total \rms to compare with the \rms of our 3D models, we computed $\mathrm{\sqrt{\sigma_1^2 + \sigma_2^2}}$, where $\sigma_1$ and $\sigma_2$ correspond to the \rms variation of two different components. For a proper comparison between the theoretical bolometric values from our 3D models and the observational measurements, which are specific to the \textit{Kepler} band pass, we have applied a bolometric correction following \citet{Ballot2011}, as this was also the approach taken by \citetalias{Kallinger2014}. We have also tried the formalism by \citet{Lund2019},  who in addition to the approach by \citet{Ballot2011} (which assumes a black-body representation of the stellar spectrum) introduced the use of synthetic 1D atmosphere models \citep[see][for further details]{Lund2019}. 

In Fig.~\ref{fig:bolcorrections} we show the standard deviation of the stars from the LEGACY sample, corrected with the two different bolometric corrections. We can see that the differences among the corrected standard deviations due to the bolometric corrections are very small, which means that they do not affect the interpretation of our results when comparing the 3D models and observations.

$\xi_i$ in Eq.~\ref{eq:bgfunc} must be determined on a star by star basis using the relation:
\begin{equation}
\xi_i = 2 a_i \sin(\pi a_i^{-1}) \, .
\end{equation}
Finally, $G(\nu)$ denotes the Gaussian envelope used to approximate the power excess from oscillations:
\begin{equation}
G(\nu) = P_g \exp\left( -\frac{(\nu - \nu_{\mathrm{max}})^2}{2\sigma_g^2} \right)\, ,
\end{equation}
where $P_g$ gives the height of the Gaussian envelope, \numax{} gives the frequency of maximum power density, and $\sigma_g$ gives the spread of the Gaussian envelope.

The background function was fitted in the interval from $3$ $\mu$Hz to the Nyquist frequency at ${\sim}8496\, \mu$Hz. We set such a lower limit on the frequency because the filtering of the data in the processing stage will inadvertently have an impact on the low-frequency part of the spectrum. The fit is performed in an MCMC manner using the \textit{emcee} Python package \citep{emcee}\footnote{\url{https://github.com/dfm/emcee}} -- we refer to \cite{Lund2017} for details on the optimization. One of the three components in \eqref{eq:bgfunc} is included to account for any low-frequency activity signal, while the other two (with the lowest timescales, $\tau$) are treated as components related to the granulation.

Based on the background fit we compute from the MCMC chains the distribution from the combined \rms signal of the two granulation-related components, by adding the respective $\sigma_i$ values in quadrature. From the resulting distribution, we determine the median and uncertainties from the limits of the $68\%$ highest probability density interval. The resulting total \rms values are provided in Table~\ref{tab:keplersample}.

Similar to the procedure adopted for the 3D simulations, we also compute a distribution for the effective e-folding timescale, \taueff, of the combination of the two granulation components by sampling from the MCMC chains of the fit. The effective timescales are provided in Table~\ref{tab:keplersample}.

\subsection{Comparison with observational data}
\label{sec:comparisonwithobservations}

In this section, we compare both \rms and \taueff obtained from our 3D stellar atmosphere models with \textit{Kepler} data, and we check if the 3D models can reproduce the observational data.

\begin{figure*}
  \centering
  \begin{subfigure}[tb]{0.49\textwidth}
    \includegraphics[width=\columnwidth]{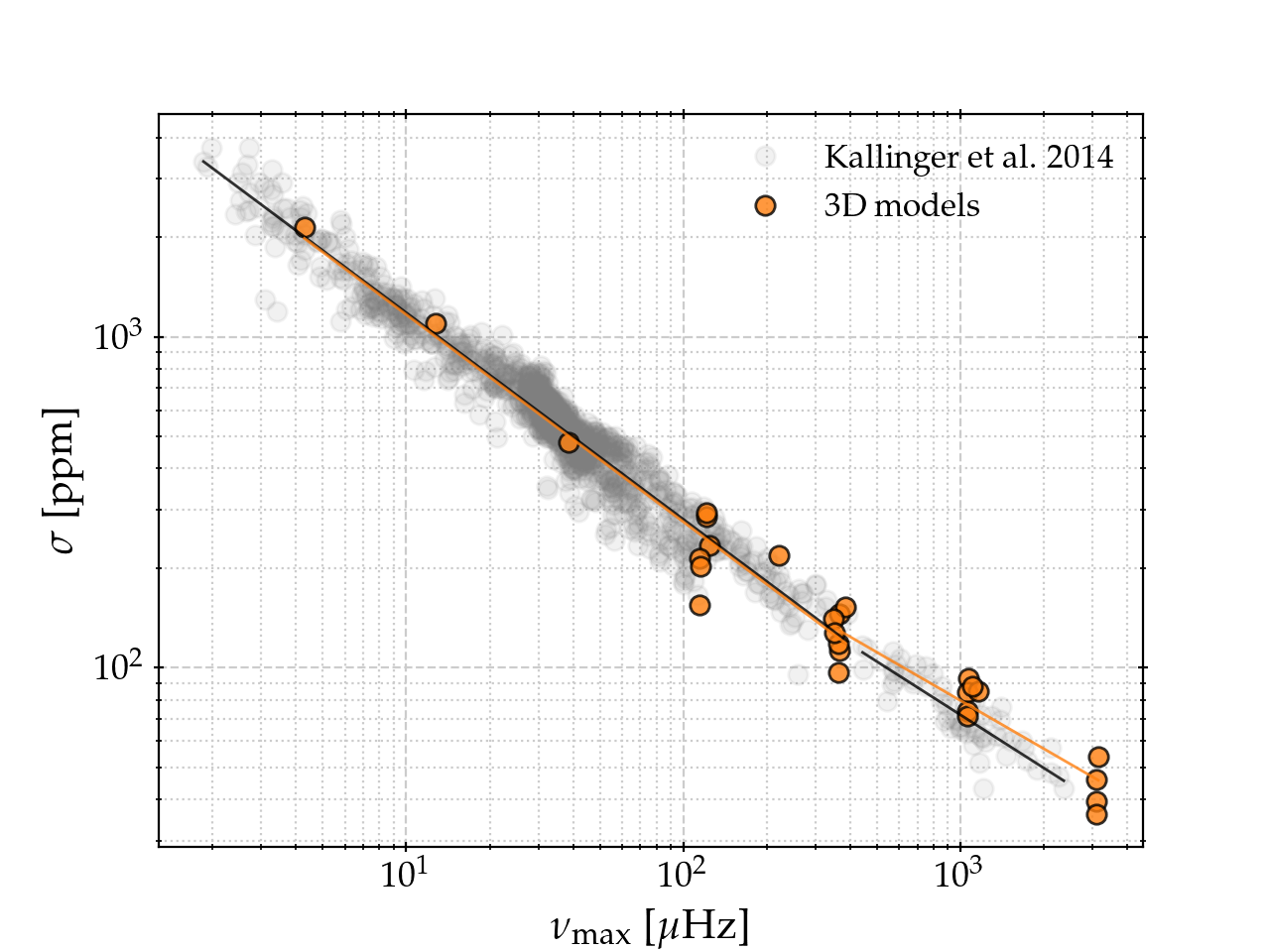}
  \end{subfigure}
  \hfill
  \begin{subfigure}[tb]{0.49\textwidth}
    \includegraphics[width=\columnwidth]{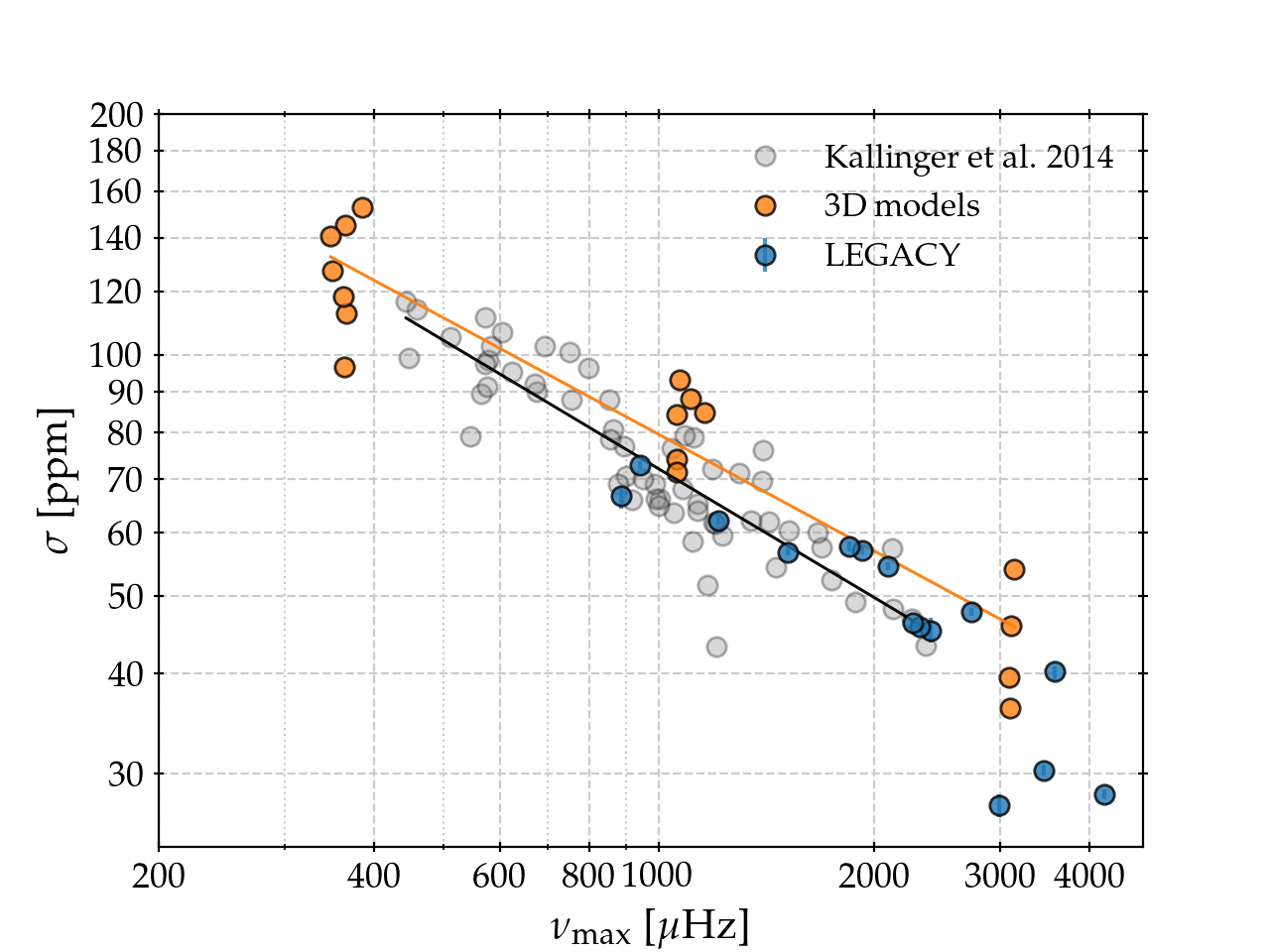}
  \end{subfigure}
  \caption{Standard deviation \rms comparison between our 3D sample and \textit{Kepler} data from \citetalias{Kallinger2014} as function of \numax. The orange lines are linear regressions of our data, while the black ones correspond to \citetalias{Kallinger2014}. For both cases, we make two kinds of fits: one for stars and models with \logg $\leq$ 3.5, and the other for \logg $\geq$ 3.5. In the right panel, we present a zoom-in comparison for \logg $\geq$ 3.5, where we added the Legacy sample.}
  \label{fig:obs-rms-numax-comparison}
\end{figure*}

We first compare the \rms values of our complete sample of 3D stellar atmosphere models --including all metallicities-- with the values reported by \citetalias{Kallinger2014}. The comparison is made in terms of \numax as shown in Fig.~\ref{fig:obs-rms-numax-comparison} and in terms of \logg in the Appendix~\ref{sec:resultslogg}. Our 3D models agree very well with the results from \citetalias{Kallinger2014}, especially for stars with \logg $\leq$ 3.5. This can be seen in the linear regressions (power laws) between our estimated \rms and \numax and \logg, which are summarized in Table~\ref{tab:RMSrelations}. The relations derived using the 3D models agree within the uncertainties with the relations from \citetalias{Kallinger2014}, but we observe a slight deviation in \rms $\propto$ $\mathrm{\nu_{max}^\alpha}$, when \logg $\geq$ 3.5.  Our 3D models tend to have a higher \rms than the ones found by \citetalias{Kallinger2014} for stars with \logg $\geq$ 4.0 or \numax $\geq$ 1000 $\mu$Hz. We preferred to have two different fits above and below \logg = 3.5 since it provides a better fit between our 3D models and observations, rather than a single one, indicating 2 possible trends in the data.

We would like to point out that including the 3D models at different metallicities results in a small scatter of \rms at a given \logg or \numax, which is also what we expected to obtain, as discussed in Sec.~\ref{sec:Resultsmetallicities}. Additionally, our models reproduce the scatter of the Kepler stars from \citetalias{Kallinger2014}.

\begin{table}
	\centering
	\caption{\textit{$\alpha$} values from the relations between \rms with $g$ and \numax from
Fig.~\ref{fig:obs-rms-numax-comparison}.}
	\label{tab:RMSrelations}
	\begin{tabular}{lcc} 
		\hline
	    Relation & \textit{$\alpha$} - 3D models & \textit{$\alpha$} - \citetalias{Kallinger2014}\\
		\hline \hline
        \multicolumn{3}{c}{ \logg $\leq$ 3.5} \\
        \hline 
	    \rms $\propto$ g$\rm{^\alpha}$& -0.609 $\pm$ 0.037 & -0.610 $\pm$ 0.004 \\
	    \rms $\propto$ $\mathrm{\nu_{max}^\alpha}$& -0.626 $\pm$ 0.041 & -0.624 $\pm$ 0.005\\
		\hline\hline
        \multicolumn{3}{c}{ \logg $\geq$ 3.5} \\
        \hline 
	    \rms $\propto$ g$\rm{^\alpha}$& -0.507 $\pm$ 0.045 & -0.529 $\pm$ 0.025 \\
	    \rms $\propto$ $\mathrm{\nu_{max}^\alpha}$& -0.510 $\pm$ 0.044 & -0.535 $\pm$ 0.026\\
	    \hline
	\end{tabular}
\end{table}

\begin{figure*}
  \centering
  \begin{subfigure}[tb]{0.49\textwidth}
    \includegraphics[width=\columnwidth]{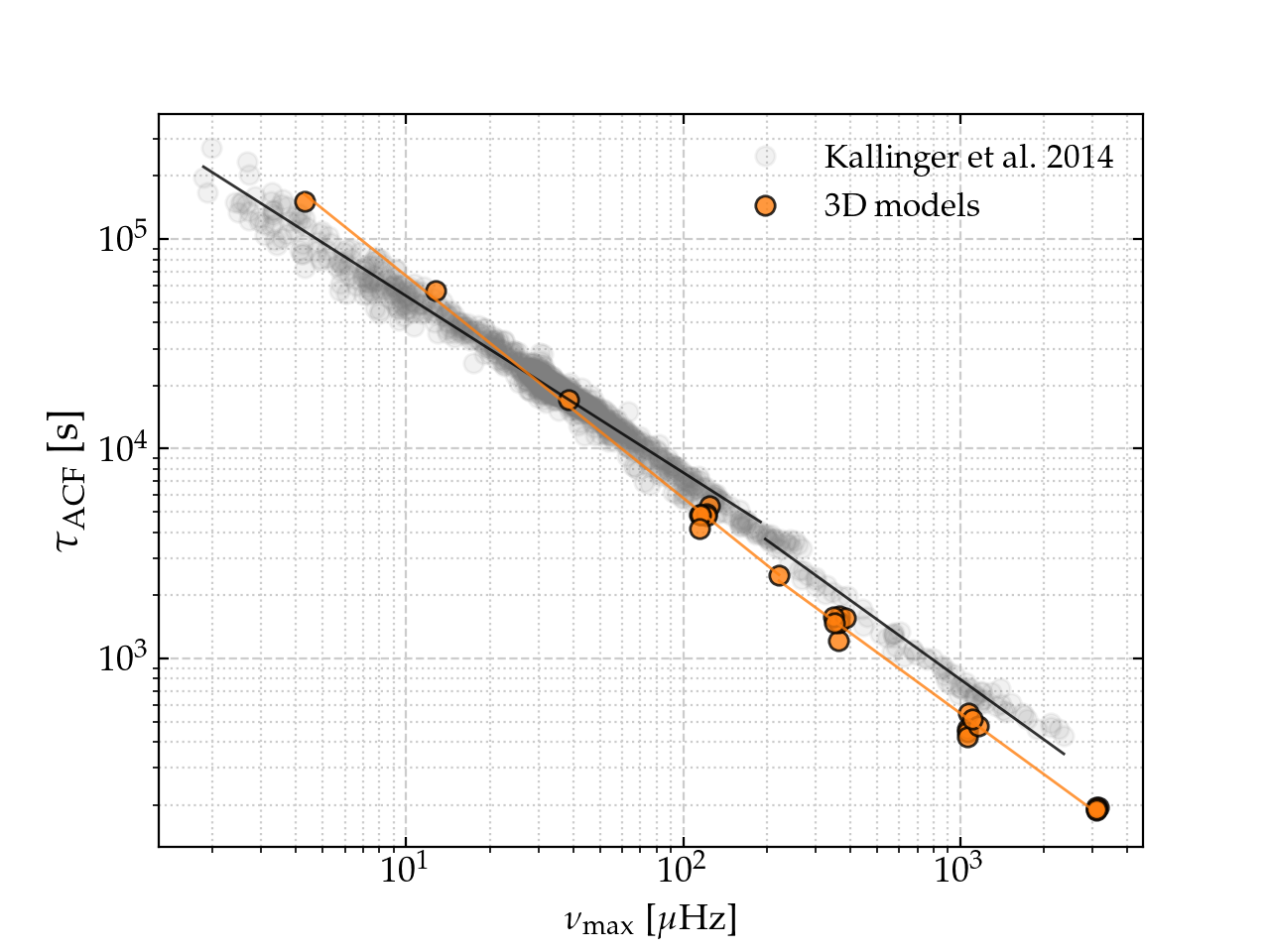}
  \end{subfigure}
  \hfill
  \begin{subfigure}[tb]{0.49\textwidth}
    \includegraphics[width=\columnwidth]{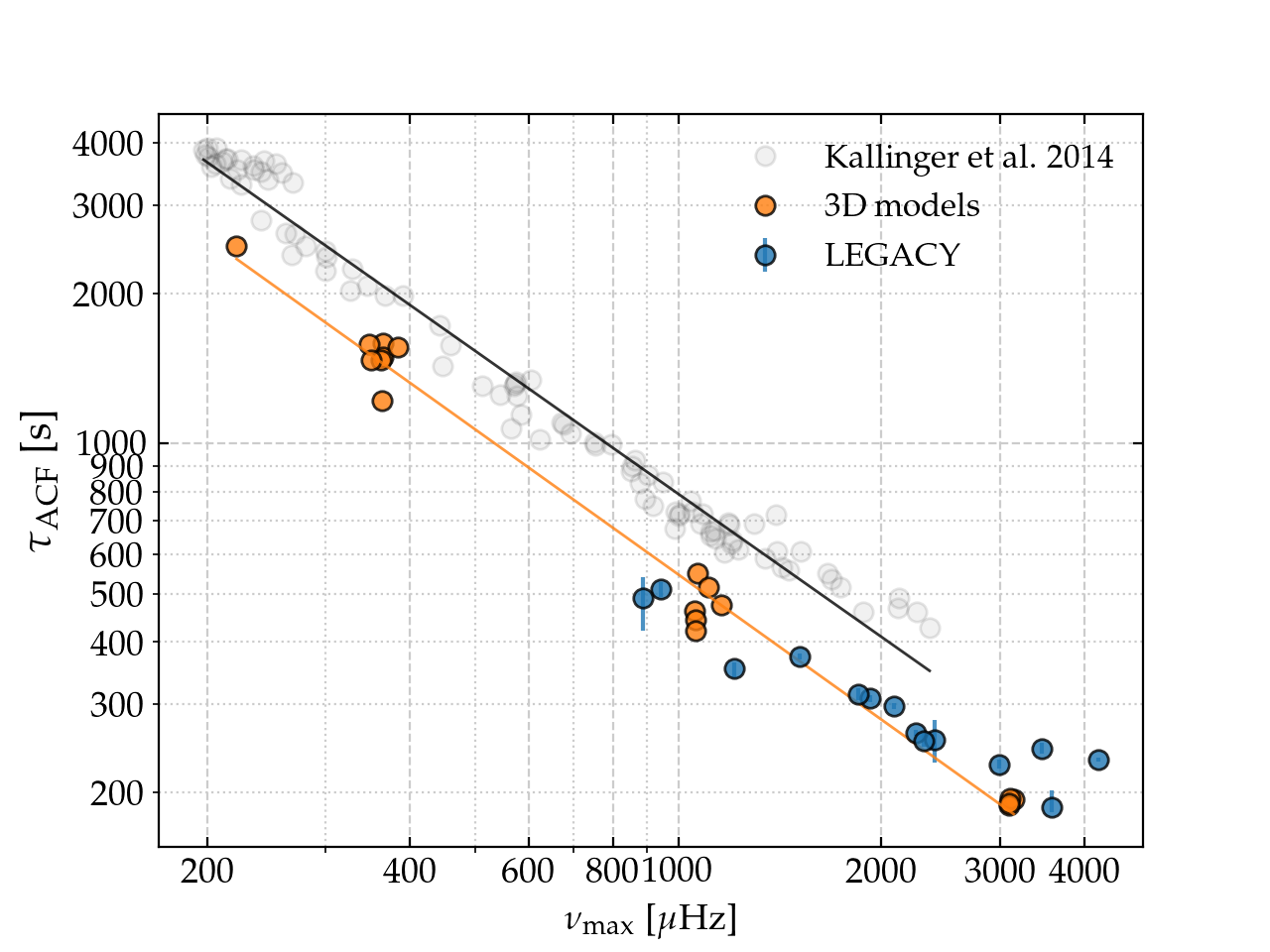}
  \end{subfigure}
  \caption{Auto-correlation time \taueff comparisons between our 3D sample and Kepler data from \citetalias{Kallinger2014}, as functions of \numax. The orange lines are linear regressions of our data, while the black ones correspond to \citetalias{Kallinger2014}. For both cases, we make two kinds of fits: one for stars and models with \logg $\leq$ 3.5, and the other for \logg $\geq$ 3.5. In the right panel, we present a zoom-in comparison for \logg $\geq$ 3.5, adding the Legacy sample. Our 3D models agree with the Legacy sample, while there is an offset with \citetalias{Kallinger2014} indicating a possible problem in stellar parameter determination.}
  \label{fig:obs-time-numax-comparison}
\end{figure*}

We decided to have a close-up analysis on the stars with \logg $\geq$ 3.5, by comparing our 3D models, \citetalias{Kallinger2014} data, and LEGACY stars. Fig.~\ref{fig:obs-time-numax-comparison} shows \rms in terms of \numax for the three data sets. We see that the LEGACY sample overlaps with some of the \citetalias{Kallinger2014} stars, but also follows our 3D models with \logg $\geq$ 4.4 or \numax $\geq$ 3000 $\mu$Hz. Some of the 3D models have slightly higher values than the ones reported by observations, but this might also be due to the sparse sample of stars. Despite that, the theoretical and observational linear regressions are very similar within the 1-sigma error bars, as shown in Table~\ref{tab:RMSrelations}. 

Regarding the auto-correlation timescales, we calculated the ACF directly using the time series from the 3D models, whereas \citetalias{Kallinger2014} used the two superposed granulation components from the fit to the power spectrum of each star. Fig.~\ref{fig:obs-time-numax-comparison} shows the comparison of our timescales from the 3D models with the ones reported by \citetalias{Kallinger2014} in terms of \numax, and in terms of \logg in the Appendix~\ref{sec:resultslogg}. As for the standard deviations, we derived two fits above and below \logg$=3.5$, which are summarized in Table~\ref{tab:Timescalesrelations}. On the one hand, the 3D models and observational timescales in general agree when \logg $\leq$ 3.5. On the other hand, the timescales from our 3D sample do not fit \citetalias{Kallinger2014} values when \logg $\geq$ 3.5, and there is a clear offset between the 3D models and data from \citetalias{Kallinger2014} of $\sim 200\mu$Hz for the sub-giants, to $\sim 500\mu$Hz for the MS stars. To further explore these discrepancies, we decided to compare with the timescales of the LEGACY sample as well, which were computed by following a similar approach as \citetalias{Kallinger2014} did, but with a more robust technique, as mentioned in Sec.~\ref{sec:RMSKepler}. 

The comparisons are illustrated in the right panel of Fig.~\ref{fig:obs-time-numax-comparison} in terms of \numax, which shows that the timescales from the LEGACY sample follow the trend depicted by the 3D models. Since the LEGACY sample included three stars also present in the sample from \citetalias{Kallinger2014}, we expected to reproduce their timescales, but that was not possible. One of the reasons behind the discrepancy between the timescales from \citetalias{Kallinger2014} and the LEGACY sample is, that the former study fixed the exponents of the background models to 4, while they were set to be free parameters in the background models of the LEGACY sample, which is important since \cite{Mathur2011} showed that different background models provide slightly different results. Another reason might be the robustness of the method used in this work to determine the timescales, compared to the one used in \citetalias{Kallinger2014}, since we computed a distribution of \taueff using MCMC. Additionally, the horizontal offset in Fig.~\ref{fig:obs-time-numax-comparison} between the LEGACY sample and data from \citetalias{Kallinger2014} might also be due to having more accurate stellar parameters derived with asteroseismology, as is the case for the LEGACY sample. Therefore, this analysis suggests that the stars analyzed in \citetalias{Kallinger2014} using scaling relations introduced some biases in the granulation power laws.

\begin{table}
	\centering
	\caption{\textit{$\alpha$} values from the relations between \taueff with \logg and \numax from 
Fig.~\ref{fig:obs-time-numax-comparison}.}
	\label{tab:Timescalesrelations}
	\begin{tabular}{lcc} 
		\hline
	    Relation & \textit{$\alpha$} - 3D models & \textit{$\alpha$} - \citetalias{Kallinger2014}\\
		\hline \hline
        \multicolumn{3}{c}{ \logg $\leq$ 3.5} \\
        \hline 
	    \taueff $\propto$ g$\rm{^\alpha}$& -1.028 $\pm$ 0.023 & -0.821 $\pm$ 0.003 \\
	    \taueff $\propto$ $\mathrm{\nu_{max}^\alpha}$& -1.060 $\pm$ 0.027 & -0.843 $\pm$ 0.004\\
		\hline\hline
        \multicolumn{3}{c}{ \logg $\geq$ 3.5} \\
        \hline 
	    \taueff $\propto$ g$\rm{^\alpha}$& -0.948 $\pm$ 0.021 & -0.901 $\pm$ 0.013 \\
	    \taueff $\propto$ $\mathrm{\nu_{max}^\alpha}$& -0.965 $\pm$ 0.026 & -0.951 $\pm$ 0.014\\
	    \hline
	\end{tabular}
\end{table}

\section{Discussion}
\label{sec:Discussion}

Several studies have determined granulation properties in terms of standard deviations and auto-correlation times (\rms, and \taueff) for stars observed by the \textit{Kepler} mission (i.e. \citet{Mathur2011}; \citetalias{Kallinger2014}; \cite{Kallinger2016, Corsaro2017}). Additionally, several attempts have been made to compare these values with 3D stellar atmosphere models (i.e. \cite{Mathur2011, Tremblay2013, Samadi2013, Ludwig2016}).

\cite{Mathur2011} analyzed the power spectra of approximately 1000 red giants with six different methods, all based on Harvey profiles, similar to for our treatment of the LEGACY sample. They found an overall agreement among the six methods since they all showed that the granulation power and \taueff increase with decreasing \numax, as we also reproduced in this study. However, the six methods did not retrieve the same values of granulation power or \taueff, due to the systematic differences associated with the choice of background fitting method. \cite{Tremblay2013} arrived to a similar conclusion by modeling the power spectrum of 3D stellar atmosphere models with different expressions, which resulted in significantly different timescales. Still, \cite{Mathur2011} reported general relations for the granulation properties, where \rms $\propto$ $\mathrm{\nu_{max}^{-0.45}}$, similar to the values we found for the 3D models with \logg$\geq$3.5.  They found an auto-correlation time \taueff $\propto$ $\mathrm{\nu_{max}^{-0.89}}$, which is an average between the timescales derived from all six methods. It shows a lower coefficient than the ones we reported in Table~\ref{tab:Timescalesrelations} for our 3D models, but their relation is similar to the $\alpha$ values found for the sample of \citetalias{Kallinger2014}.

\cite{Mathur2011} compared as well their values with 3D models of red giants at solar metallicities, which belong to the grid described in \cite{Trampedach2013}. They reported that both the effective timescales and granulation power of the simulations decrease as \numax increases, as shown in this study. However, their timescales were overestimated and granulation power laws were underestimated when compared with observational data, which is not the case in our study, since we reproduce both the observed \rms and timescales. They argued that the discrepancies might be due to only having models at solar metallicity, or stellar radii with insufficient accuracy, among other explanations. Based on our results, we do acknowledge that metallicity has an impact on \rms, hence the granulation power, and \taueff, but it cannot cause a systematic shift of these quantities as shown in Fig. 11 in \cite{Mathur2011}. Moreover, as shown in Fig.~\ref{fig:CompMetallicities}, for a given set of \teff-\logg, the quantities \rms and \taueff decrease slowly with metallicity and cannot compensate the discrepancies reported by them. Similarly, we point out that, even though inaccurate stellar radii associated with the 3D models do affect the standard deviation, the results should be more scattered rather than systematically shifted. Thus, we attribute the discrepancies to the short time series of the 3D models used in \cite{Mathur2011}, and possibly to the presence of box modes  (see Appendix~\ref{sec:effects}).

We also compared our results with \cite{Samadi2013} as shown in Fig.~\ref{fig:CompSamadi}. They analyzed their 3D models by modeling the power density spectrum and derived the \rms brightness fluctuations from it. However, their \rms values have strong scatter if expressed only in terms of \numax, while the scatter of \taueff is less significant. The scatter is so critical, that for models with \numax above 1000$\mu$Hz, they predicted that \rms can oscillate 80 ppm for a given \numax and, in some cases, fluctuate between two orders of magnitude. 

To find a better correlation with stellar parameters, they proposed new scaling relations for both quantities with a dependency on the Mach number, i.e. the ratio between the convective velocity and the sound speed. Thus, it is a difficult quantity to obtain for observed stars. On the contrary, our 3D models and their standard deviation of the brigthness fluctuations clearly follow a simple power law in terms of \numax. Hence, we show that it is not necessary to include the Mach number dependence in the scaling relations.  Even though we do not explore the relationship with the Mach number, we refer the reader to \cite{Tremblay2013} for a discussion about the granulation properties in terms of the Mach number.

\citetalias{Kallinger2014} expanded the study made by \cite{Mathur2011} by including MS stars, and tested several methods to model the granulation background signal using a probabilistic approach. 
They originally reported that \rms $\propto$ $\mathrm{\nu_{max}^{-0.61}}$, but derived as well a relation dependant on the stellar mass to follow the 3D models from \cite{Samadi2013}. However, we do want to point out that we observed the same behavior in the LEGACY stars analyzed in this sample, so the change of trend is unlikely to be associated with the method used to analyze both 3D models and stars. Future studies that decide to explore this problem should focus on adding more MS stars at different metallicities with accurate stellar parameters. Additionally, more 3D models with, for instance, \numax = 600 $\mu$Hz or 2000 $\mu$Hz should be added to fill in the gaps shown in Fig.~\ref{fig:obs-rms-numax-comparison}. 
Regarding the timescales, \taueff has a different behavior than the relation derived with observational data from \citetalias{Kallinger2014} for \logg $\leq$ 3.5. For the case \logg $\geq$ 3.5 we see that the theoretical and observational relations are more similar, but still not the same. Additionally, there seems to be an offset between the 3D models and LEGACY sample with \citetalias{Kallinger2014}. These discrepancies might arise due to the accuracy of the stellar parameters, or the different methods used to model the power spectrum density, as \cite{Mathur2011} found out. Thus, it seems that the function to fit the power spectrum density has a great impact on the derived timescale, as \cite{Tremblay2013} suggested as well. Since the LEGACY sample reproduces the timescales from our 3D models, we propose that the fits to the power spectrum of stars keep the exponent as a free parameter, contrary to what \citetalias{Kallinger2014} did.   

We attempted to derive the timescales as \cite{Kallinger2016} proposed, that is, by fitting a \textit{sinc$^2$} function to the ACF of filtered time series of \textit{Kepler} stars, which had the combined contribution of granulation and p-mode oscillations. We note that they did not actually include the factor of four in the fit proposed in the Supplementary Materials  \citep{Kallinger2016} in their calculations, so we did not use it either. However, we could not reproduce the observed timescales, since our values were almost overestimated by a factor of two. One of the reasons might be due to the p-mode contribution from the observational data since our 3D models did not have any contribution from oscillations. 

Lastly, we comment on the effect of metallicity on \rms. To our knowledge, \cite{Ludwig2016} is the only previous theoretical study focused on that problem, using 3D stellar atmosphere models from the CIFIST grid with metallicities equal to [Fe/H] = 0.0 and -2.0. Even though their \rms also increases with decreasing \numax, their data are scattered for both metallicities, similarly as the data presented in \cite{Samadi2013}. This is not surprising, since they also used CIFIST 3D models, but employed a different method to determine \rms. Therefore, the dispersion on the data might not be attributed to the method, but possibly to other factors, such as the length of the time series, or the radius determination of a given simulation, since it is a difficult quantity to compute by only providing atmospheric parameters.
Moreover, \cite{Ludwig2016} also claimed that they could not provide a fit for \rms in terms of [Fe/H] using their models at [Fe/H] = 0.0 and [Fe/H] = -2.0 simultaneously, and neither could reproduce the \rms values from \citetalias{Kallinger2014}, at least using the 3D  models at solar metallicity. However, in Fig.~\ref{fig:obs-rms-numax-comparison} we showed that we can provide a fit to all our 3D models together and we can reproduce the results from \citetalias{Kallinger2014} (see
Sec.~\ref{sec:comparisonwithobservations}). 

\section{Conclusions}
\label{sec:conclusions}
Thanks to photometric data from space missions, the study and characterization of the stochastic brightness fluctuations of stars has been possible. In this study we aimed to study in detail the granulation properties of these stochastic brightness fluctuations, by using 3D stellar atmosphere models distributed across the HR diagram, at different metallicities. Particularly, the granulation properties that we determined were the standard deviation of the granulation noise \rms and the characteristic timescale \taueff. For these 3D models, we computed very long time series that covered at least 1000 turnover convective times, in order to have a standard deviation independent of the length of the time series.
Unlike previous studies, and for the first time, we show that both standard deviation and auto-correlation time follow some power laws in terms of only \numax or \logg. This means that we only need to determine accurate stellar parameters to derive two properties associated to granulation. For the solar metallicity case, we found that \rms $\propto$ $\nu\rm_{max}^{-0.567}$ and \taueff $\propto$ $\nu\rm_{max}^{-0.997}$. This shows that, we did not need to introduce the Mach number in our power laws to describe the dependence of granulation properties on stellar parameters, which is a difficult quantity to determine from stellar observational constraints, since some assumptions have to be made in order to express the Mach number in terms of the fundamental parameters (Teff, logg, [Fe/H]). However, the present work shows that the introduction of this quantity is not needed to reproduce the observed granulation properties. Moreover, we found that both \rms and \taueff decrease with metallicity with a power law dependence in terms of \numax, that we derived for [Fe/H] = 0.0, -0.5, -1.0, -2.0, which means that metallicity does need to be taken into account when deriving granulation properties.

To validate our theoretical results and power laws, we compared with a large sample of \textit{Kepler} stars from \citetalias{Kallinger2014}, and selected stars from the LEGACY sample \citep{Lund2017},  since they have accurate stellar parameters. Unlike previous studies, we found that our 3D stellar atmosphere models agree very well with the granulation noise from \textit{Kepler} stars, especially at \logg $\leq$ 3.5. We think that the comparison between stars and 3D models for \logg $\geq$ 3.5 can be improved if more targets are added in that regime. Regarding the relationship between \taueff and \numax, we found that 3D models reproduced the timescales of the selected stars of the LEGACY sample, but did not fully reproduced the timescales from the \citetalias{Kallinger2014} sample. We think that this is because the stars from the LEGACY sample have accurate stellar parameters i.e., accurate \numax, while the \numax values of the stars from the \citetalias{Kallinger2014} sample were derived with asteroseismic scaling relations. 

With this study we intended to provide better scaling relations to characterize the convective granulation noise, so that we can better understand the signature of granulation in the time series of stellar targets, and hence, improve the detection and characterization of exoplanets. In a future work, we plan to expand this study by performing a full monochromatic radiative transfer within the Kepler, or PLATO bandwidths, including the response functions of these instruments.   

\section*{Acknowledgements}
Funding for the Stellar Astrophysics Centre is provided by The Danish National Research Foundation (Grant agreement no.: DNRF106). L. F. R. D. and V. A. B-K. acknowledges support from the Independent Research Fund Denmark (Research grant 7027-00096B) and the Carlsberg foundation  (grant agreement CF19-0649).
The numerical results presented in this work were obtained at the Centre for Scientific Computing, Aarhus \url{http://phys.au.dk/forskning/cscaa/}. Part of the calculations have also been performed using the OCA/SIGAMM mesocentre. This research was undertaken with the assistance of resources provided at the NCI National Facility systems at the Australian National University through the National Computational Merit Allocation Scheme supported by the Australian Government. This work was supported by the “Programme National de Physique Stellaire” (PNPS) of CNRS/INSU co-funded by CEA and CNES."

\section*{Data Availability}
Data available on request. The data underlying this article will be shared on reasonable request to the corresponding author.



\bibliographystyle{mnras}
\bibliography{example} 



\appendix
\section{Effects of the simulation setup}
\label{sec:effects}
The brightness fluctuations are intimately linked to radiative losses at the surface. Therefore the  numerical treatment of radiative transport in the RHD code might affect our results. In this section we summarize the impact that different setups have on the brightness fluctuations from the 3D stellar atmosphere models. Specifically, we analyzed the number of bins included in the opacity tables, the number of rays used to solve the Radiative Transfer (RT) equation, the spatial grid resolution and, the presence of box-modes. 

\subsection{Number of opacity bins}
\label{appendix:numberofbins}
The number of bins used in the generation of an opacity table determines the accuracy of the radiative heating and cooling rates of each 3D stellar model. This means, that both the temperature stratification and the temperature fluctuations of every model are affected by the choice of number of bins, and how they are selected. Therefore, to test the effect of the number of bins on the brightness fluctuations, two opacity tables, with 6 and 12 bins respectively, were used to create two solar models with the same setup. The time series generated for every model cover approximately 80 hours.

\begin{figure}
\includegraphics[width=\columnwidth]{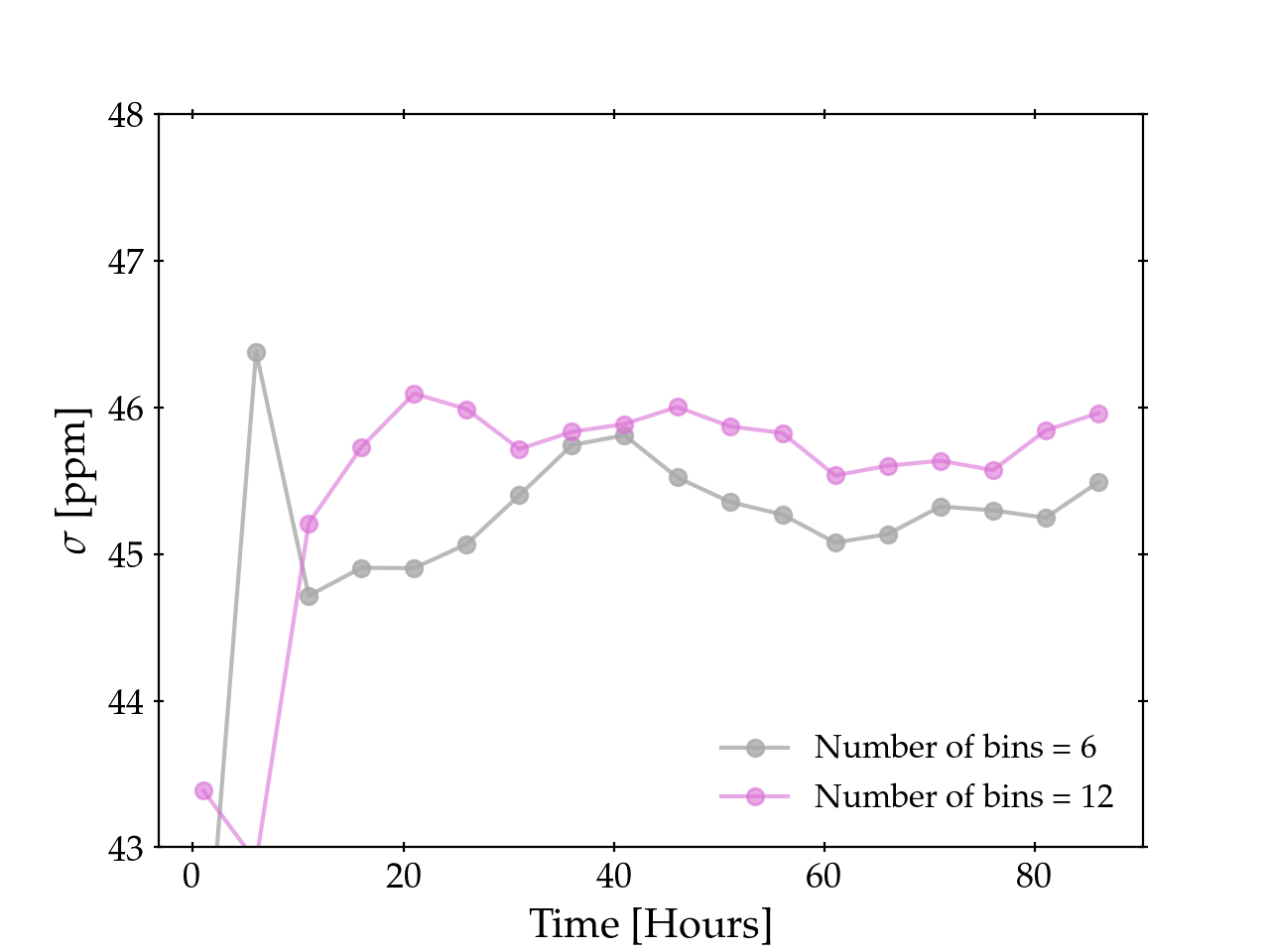}
    \caption{Effect of a 6-bin and a 12-bin opacity tables on the \rms calculated from the relative flux variations of two 3D solar models, computed with the same setup.}
    \label{fig:test_bins}
\end{figure}

Fig.~\ref{fig:test_bins} shows the \rms of brightness fluctuations calculated from the relative flux variations. The \rms was calculated with cumulative intervals of 5 hours as well.

The initial \rms values fluctuate significantly in both solar simulations as a result of the convection itself, and not as a consequence of the bin schemes in each opacity table. The \rms of both simulations evolve in a similar fashion and take very similar values. Consequently, we can conclude that the number of bins will not affect the \rms associated with the relative flux variations, as \cite{Tremblay2013} also suggested. Therefore, we decided to compute our 3D stellar models using a 6-bin opacity table, since it is less computationally expensive

\subsection{Number of rays used in radiative transfer}
\label{sec:numberrays}
In the STAGGER-code, the radiative transfer equation can be solved along a vertical ray or a combination of vertical plus inclined rays. When only a vertical ray is used, the stratifications of the thermodynamic variables are usually overestimated, including the effective temperature of the stars. This occurs because the stellar model assumes that all the radiation coming from the bottom boundary of the simulation travels vertically, which is not a realistic approach. On the other hand, the combination of one vertical ray plus one inclined, or one vertical and two inclined produce more accurate representations of the stellar atmospheres.

\begin{figure}
\includegraphics[width=\columnwidth]{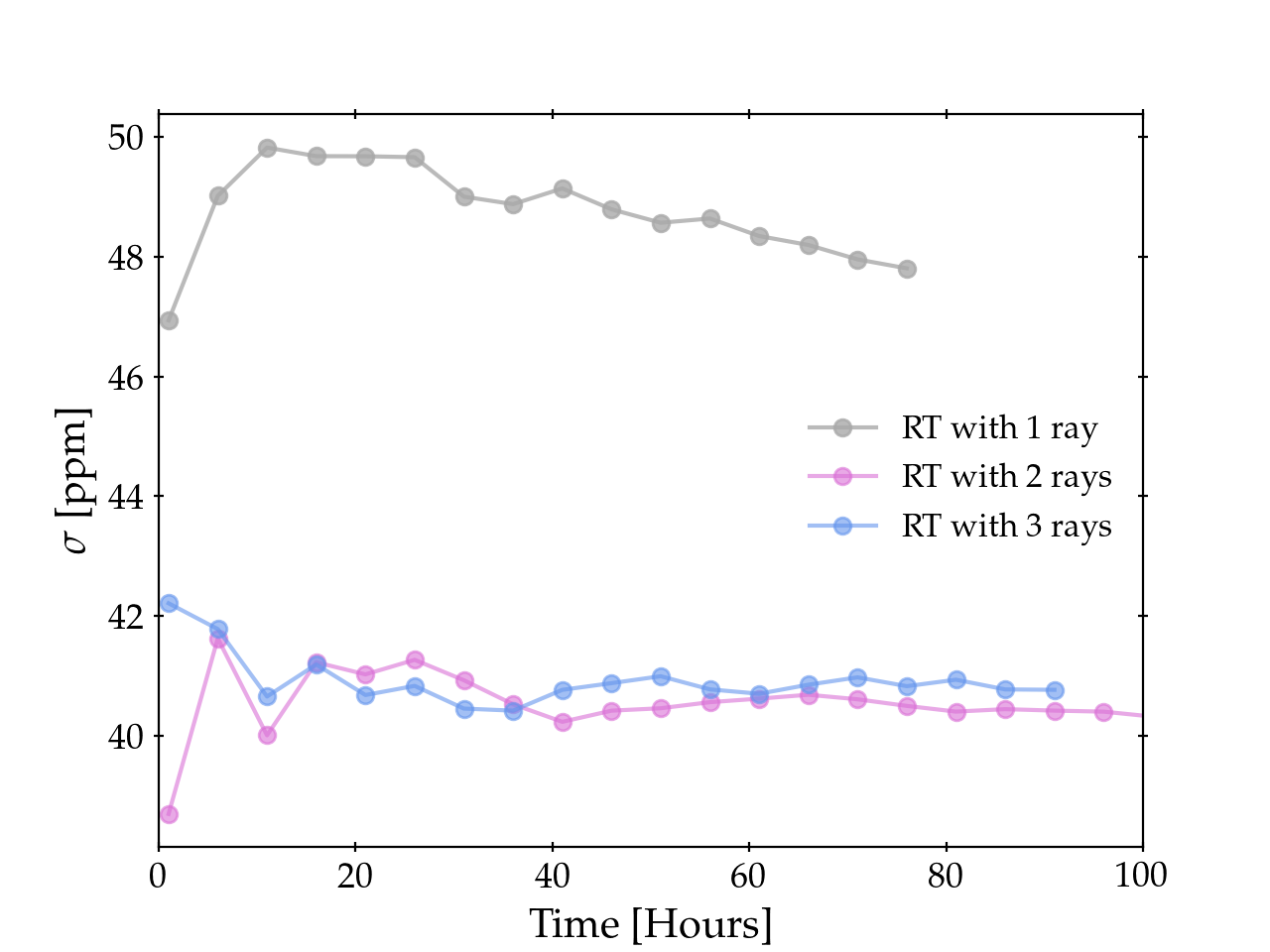}
    \caption{Effect of the number of rays used to solve the Radiative Transfer equation on the \rms of brightness fluctuations. This was calculated from the relative flux variations of three 3D solar models with identical setup. The gray series shows the one-vertical ray case, the pink series illustrates the case with two rays: one vertical plus one inclined. Finally, the blue series represents the case with three rays: one vertical plus two inclined.}
    \label{fig:test_rays}
\end{figure}

In order to test the response of the brightness fluctuations based on the number of rays, three solar simulations were generated with the three combinations of rays previously mentioned. The results obtained are shown in Fig.~\ref{fig:test_rays}, where an overestimation of the \rms is obtained when only one vertical ray is used. This is a consequence of the overestimation of the radiative flux, as it was mentioned before. On the contrary, the cases including at least one inclined ray show that \rms converges to the same value after 20 hours. In other words, using two or three rays to solve the radiative transfer equation results in basically the same \rms. Since creating a 3D stellar model is less computationally expensive if only two rays -- one inclined and one vertical -- are used, we decided to compute our sample of 3D models with that configuration.

\subsection{Effect of grid resolution and box-modes}
The 3D stellar models computed with the STAGGER-code have typically the following grid resolutions: low resolution, where every direction has 120 points; and high resolution, where each has 240 points. 
Furthermore, the simulations also include box-modes \cite{Nordlund2001}, which are acoustic modes generated by both the stochastic excitation and the work done by the bottom boundary. These modes have, however, excessively large amplitudes, due to the shallow nature of the simulation boxes and, consequently, due to their very low inertia.

In order to assess the possible impacts of both the resolution and box modes, four solar models were created as follows:

\begin{enumerate}
    \item Damping of box modes and low resolution (120$^3$).
    \item Damping of box modes and high resolution (240$^3$).
    \item No damping of box modes and low resolution (120$^3$).
    \item No damping of box modes and high resolution (240$^3$).
\end{enumerate}

\begin{figure}
\includegraphics[width=\columnwidth]{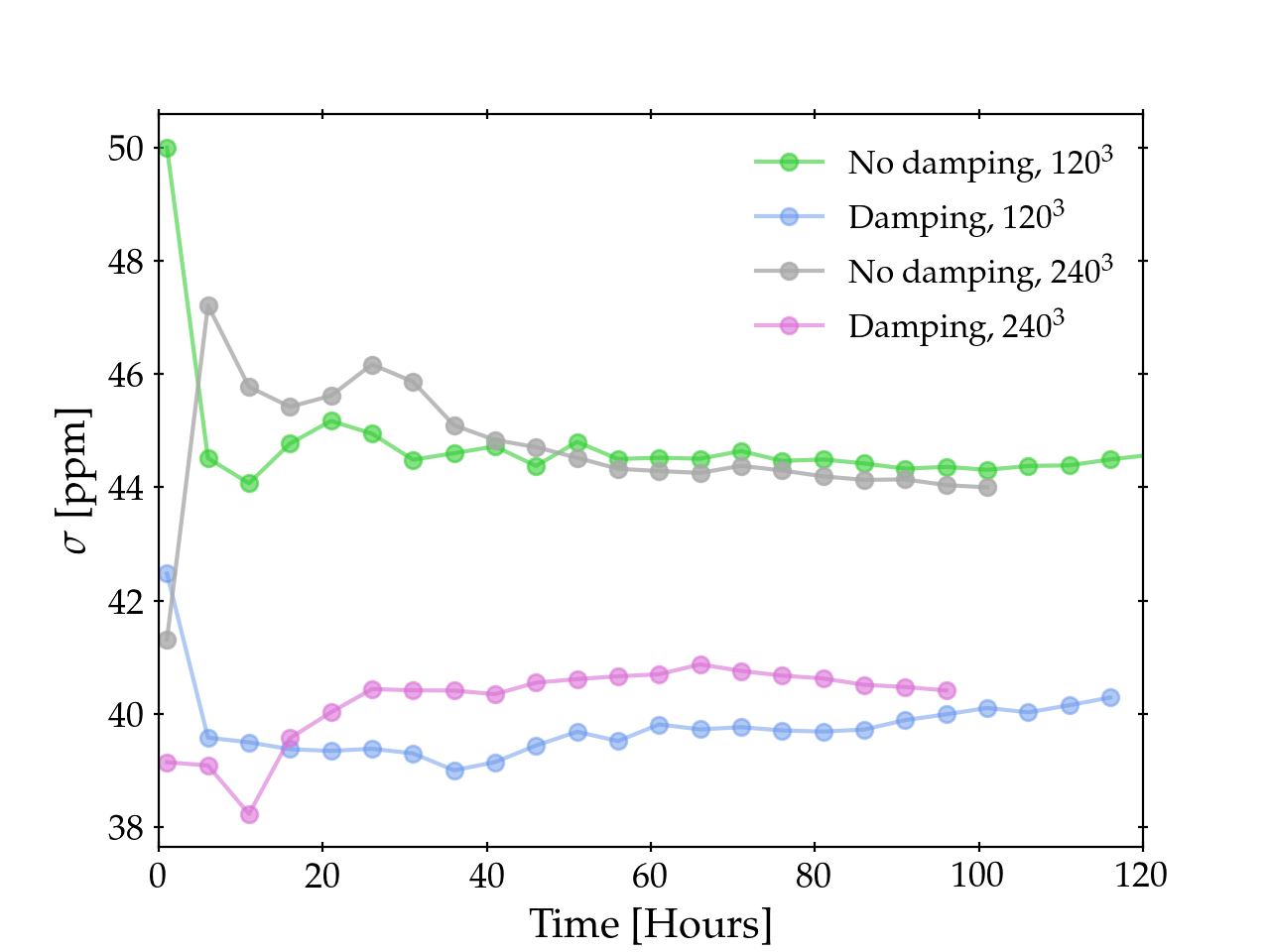}
    \caption{Effect of the resolution and box modes on the \rms calculated from the relative flux variations of 3D solar models, computed with the same setup. }
    \label{fig:test_res}
\end{figure}

The \rms from these 3D models are shown in Fig.~\ref{fig:test_res}. It can be seen that the resolution does not have a significant influence on the results, if the time series are long enough, which is the case for our models. Thus, we decided to compute the models listed in Table~\ref{tab:sample} in the low resolution regime, for computational purposes.  

On the other hand, the box modes do have a significant impact on \rms, especially in the low resolution models. However, since the large box mode amplitudes of the simulations overestimate the observed modes in stars, we think that damping the box modes provides a more reliable approach for this study.  

\section{\texorpdfstring{$\sigma$}{rms} of the relative brightness fluctuations}

\begin{figure}
	\includegraphics[scale=0.5]{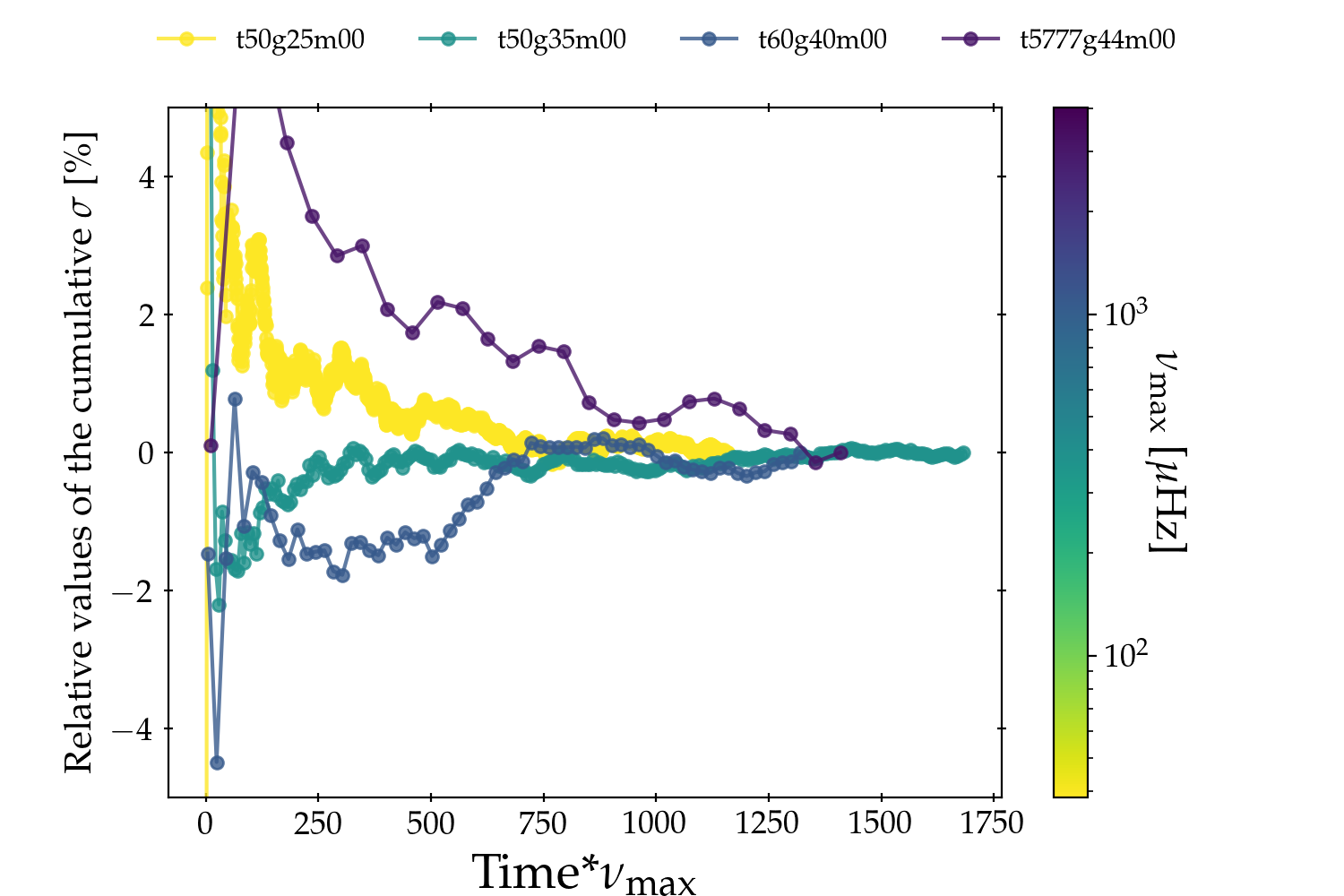}
    \caption{Relative \rms for four different 3D stellar atmosphere models at solar metallicity, as a function of the number of convective turnover times.}
    \label{fig:relativeRMSsample}
\end{figure}

Fig.~\ref{fig:relativeRMSsample} shows the convergence of \rms on four different 3D stellar atmosphere models, and the importance of having very long time series. 

\section{Comparison of standard deviation in terms of surface gravity}
\label{sec:resultslogg}
\begin{figure*}
  \centering
  \begin{subfigure}[tb]{0.49\textwidth}
    \includegraphics[width=\columnwidth]{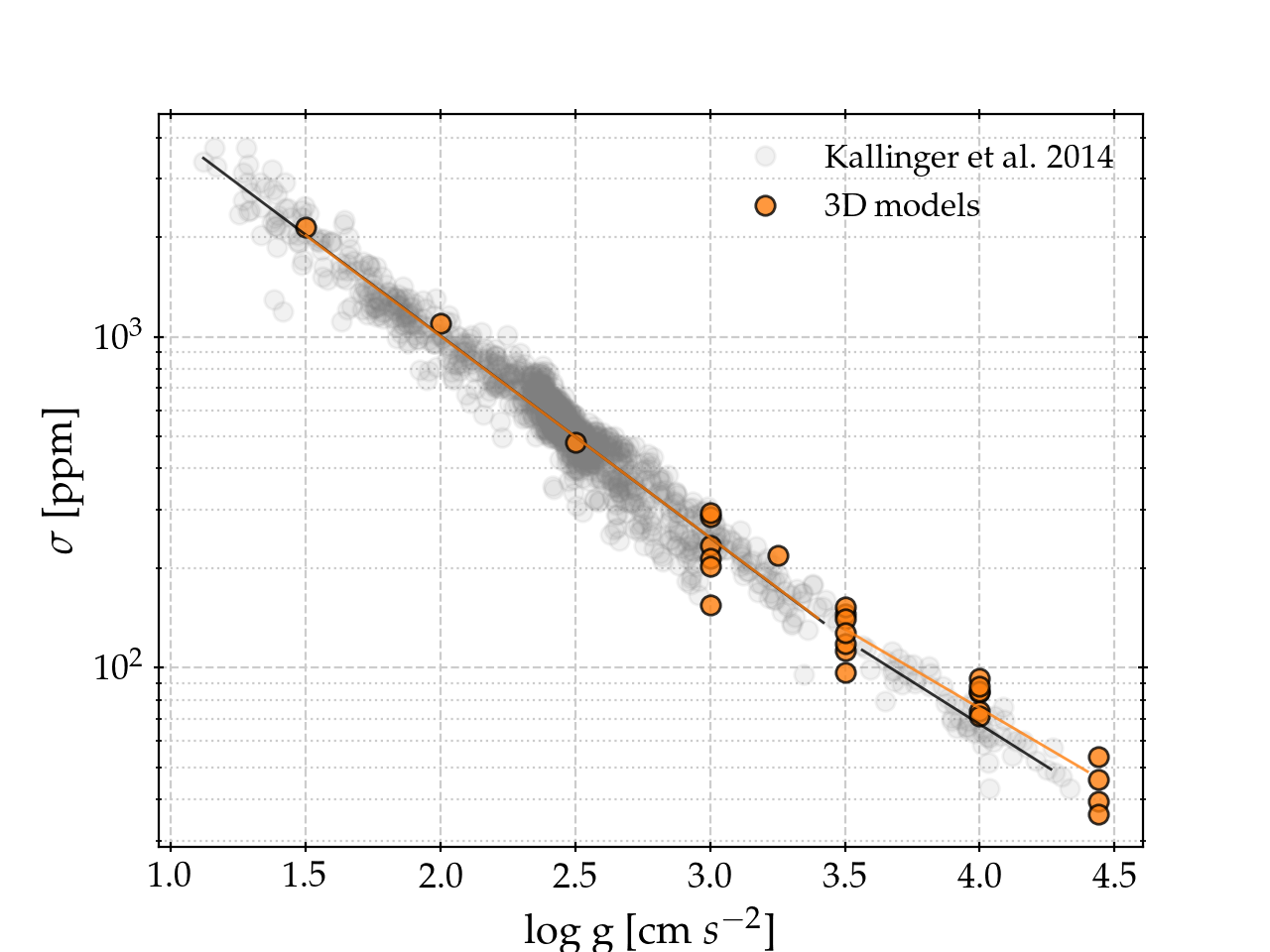}
  \end{subfigure}
  \hfill
  \begin{subfigure}[tb]{0.49\textwidth}
    \includegraphics[width=\columnwidth]{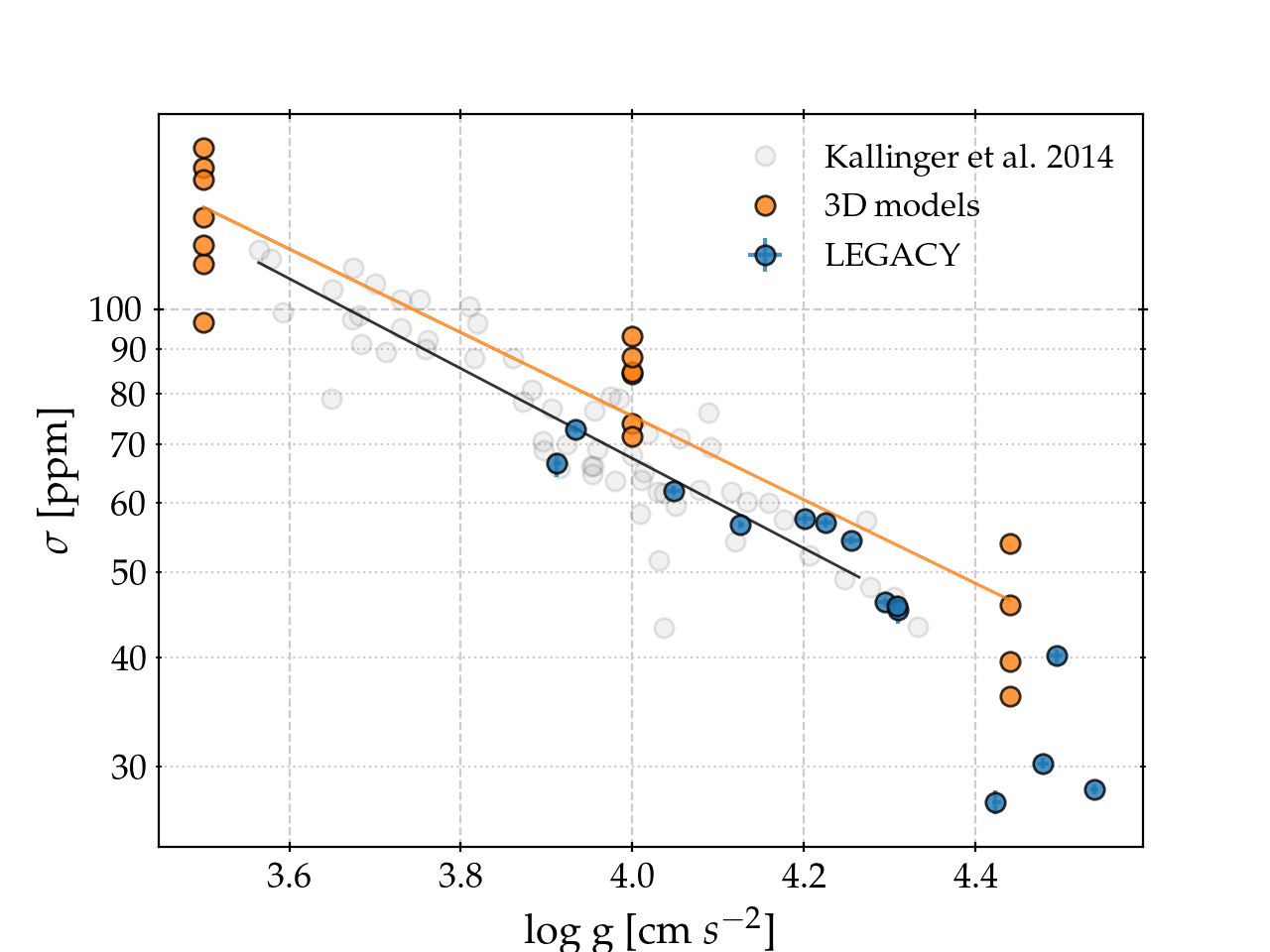}
  \end{subfigure}
  \caption{Similar to Fig.~\ref{fig:obs-rms-numax-comparison}, where we show the comparison between the standard deviation \rms of our 3D sample and \textit{Kepler} data from \citetalias{Kallinger2014} as function of \logg.  The orange lines are linear regressions of our data, while the black ones correspond to \citetalias{Kallinger2014}. For both cases, we make two kinds of fits: one for stars and models with \logg $\leq$ 3.5, and the other for \logg $\geq$ 3.5. In the right panel, we present a zoom-in comparison for \logg $\geq$ 3.5. adding the Legacy sample. In Table~\ref{tab:RMSrelations} we summarize the values for the the relation in between \rms and $g$.}
  \label{fig:rms-comparison-obs-logg}
\end{figure*}

\begin{figure*}
  \centering
  \begin{subfigure}[tb]{0.49\textwidth}
    \includegraphics[width=\columnwidth]{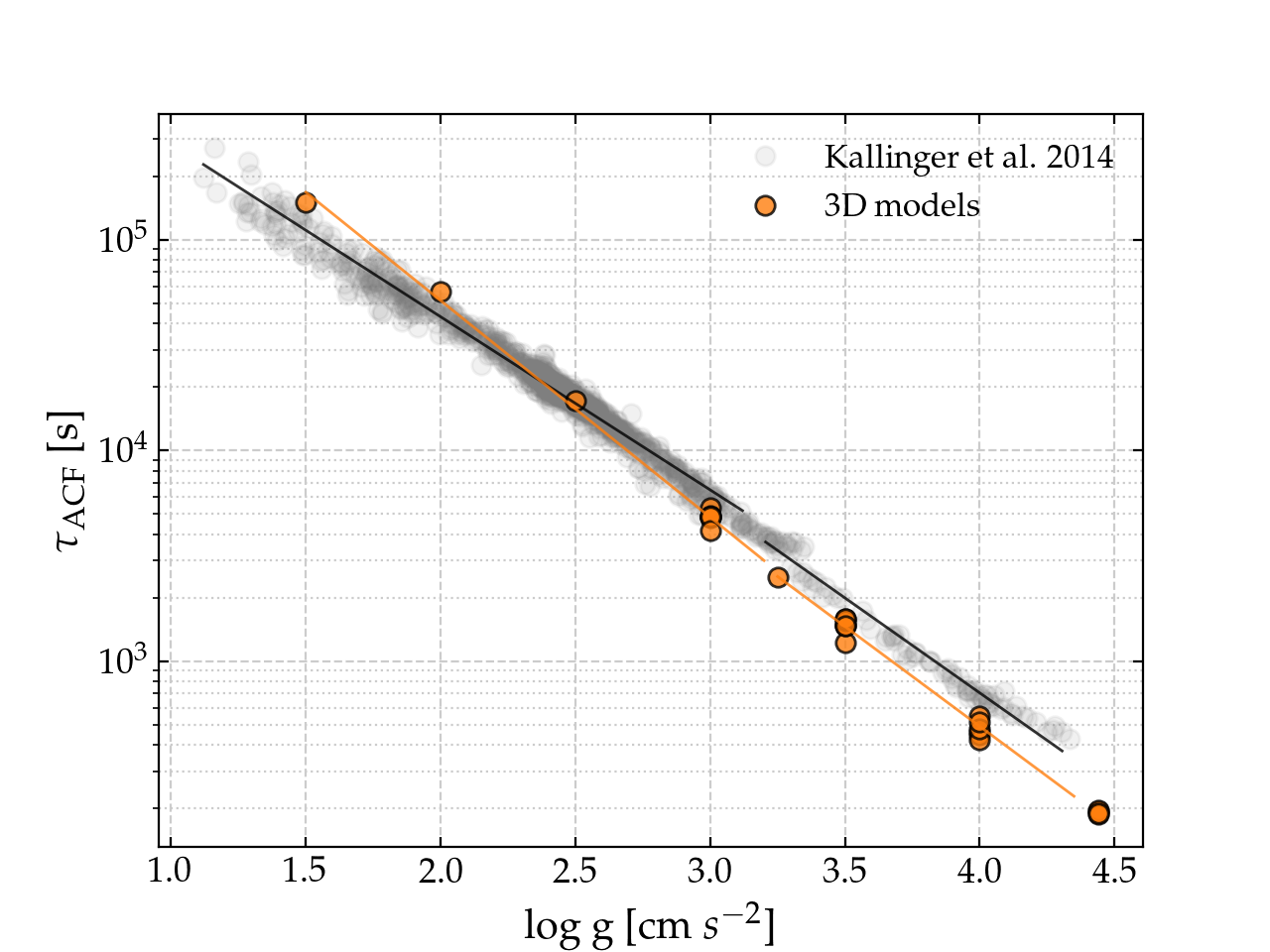}
  \end{subfigure}
  \hfill
  \begin{subfigure}[tb]{0.49\textwidth}
    \includegraphics[width=\columnwidth]{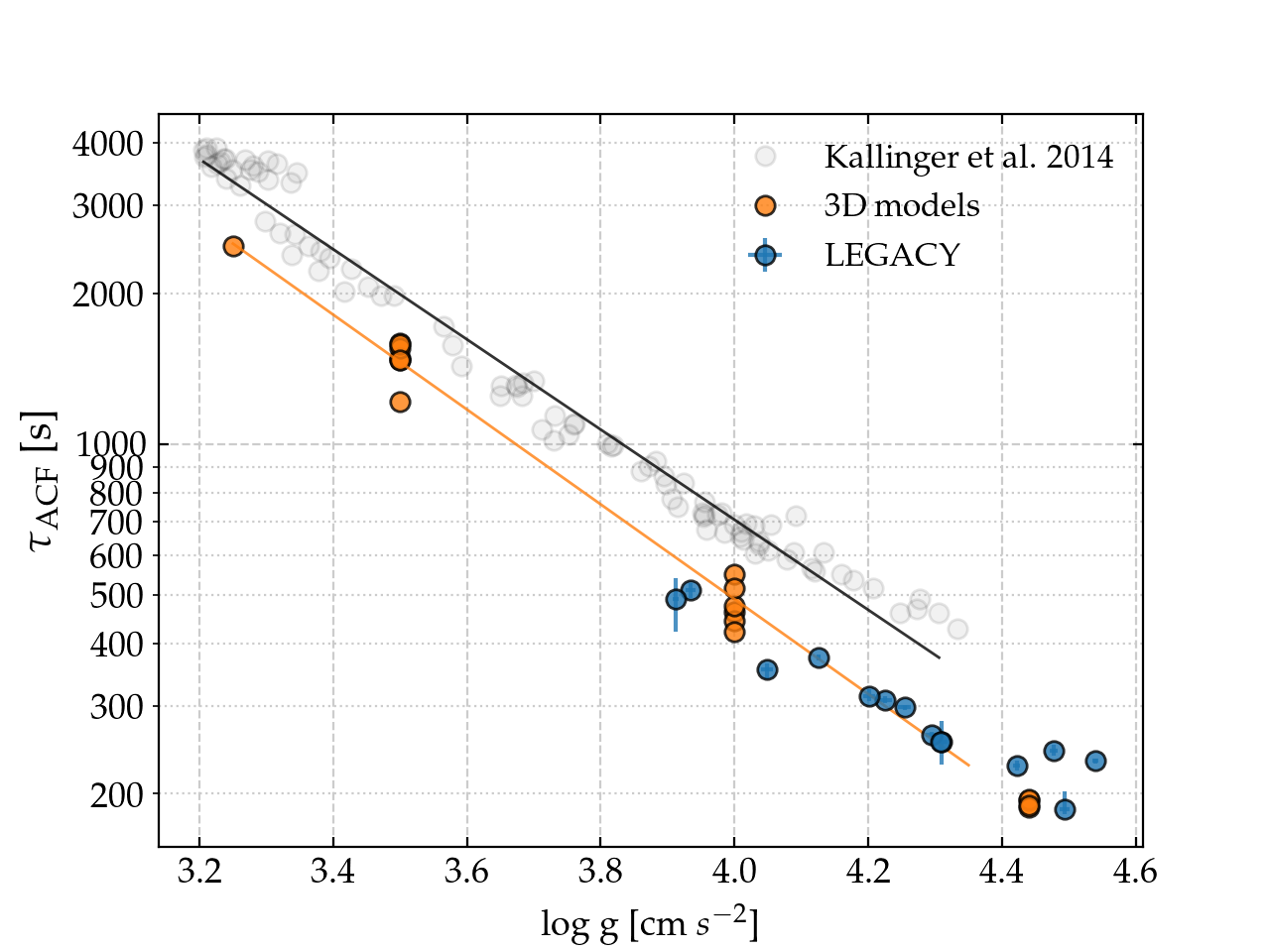}
  \end{subfigure}
  \caption{Similar to Fig.~\ref{fig:obs-time-numax-comparison}, where we show the comparison between the auto-correlation time \taueff of our 3D sample and \textit{Kepler} data from \citetalias{Kallinger2014} as function of \logg.  The orange lines are linear regressions of our data, while the black ones correspond to \citetalias{Kallinger2014}. For both cases, we make two kinds of fits: one for stars and models with \logg $\leq$ 3.5, and the other for \logg $\geq$ 3.5. In the right panel, we present a zoom-in comparison for \logg $\geq$ 3.5. adding the Legacy sample. In Table~\ref{tab:Timescalesrelations} we summarize the values for the the relation in between \rms and $g$.}
  \label{fig:time-comparison-obs-logg}
\end{figure*}

In this section we show the standard deviation \rms and the auto-correlation timescales \taueff in terms of surface gravity \logg. Fig.~\ref{fig:rms-comparison-obs-logg} shows the comparison of the standard deviation with surface gravity, while Fig.~\ref{fig:time-comparison-obs-logg} shows the auto-correlation timescales in terms of surface gravity.

\section{Effect of bolometric correction on the standard deviation of the LEGACY sample}
\label{sec:bolcorrection}

As mentioned in Sec.~\ref{sec:RMSKepler}, we decided to test the effect of the bolometric correction just on the LEGACY sample. We compared the formalism proposed by \citet{Ballot2011}, where $C\rm_{bol} = (T\rm_{eff}/T\rm_0)^\alpha$ with $T\rm_0 = 5934 \ K$ and $\alpha = 0.8$, with the formalism by \citet{Lund2019}, who used two methods to calculate the bolometric correction. The first method is assuming a stellar black-body (Planck) spectrum, similar to \citet{Ballot2011}, while the second method uses 1D atmosphere models \citep[see][for further details]{Lund2019}.

\begin{figure}
	\includegraphics[width=\columnwidth]{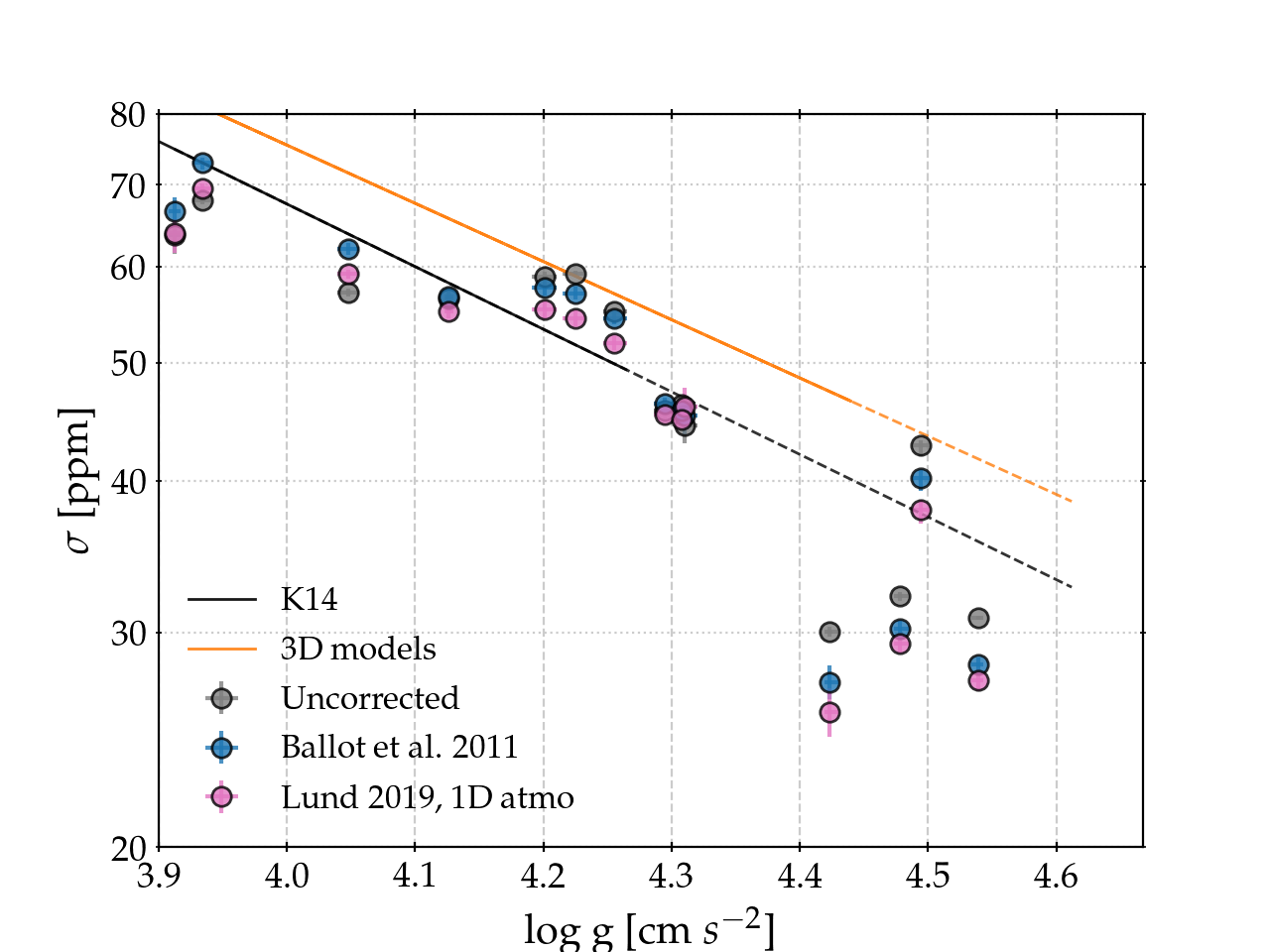}
    \caption{Comparison of the standard deviation values for the
LEGACY sample, corrected using the two different approaches to calculating the bolometric
correction. In grey, we show the uncorrected values; in blue the ones corrected the formalism by \citet{Ballot2011}, corresponding to the values used in our work; and in pink the ones corrected using the synthetic 1D atmosphere approach by \citet{Lund2019}. For comparison, we have included the linear regression relations from our 3D sample (orange line) and the corresponding from \citetalias{Kallinger2014} (black line). We extrapolated both relations for visualization, which are shown as dashed lines.}
    \label{fig:bolcorrections}
\end{figure}

In Fig.~\ref{fig:bolcorrections} we show the standard deviation of the stars from the LEGACY sample, corrected with the three different bolometric corrections. We can see that the differences among the corrected standard deviations due to the bolometric corrections are very small --less than $5 \ ppm$ in all targets--, which means that they do not affect the interpretation of our results when comparing the 3D models and observations.

\section{Power spectra fitting for the selected stars of the LEGACY sample}
\label{sec:psdfitting}
\begin{figure*}
	\includegraphics[scale = 0.5]{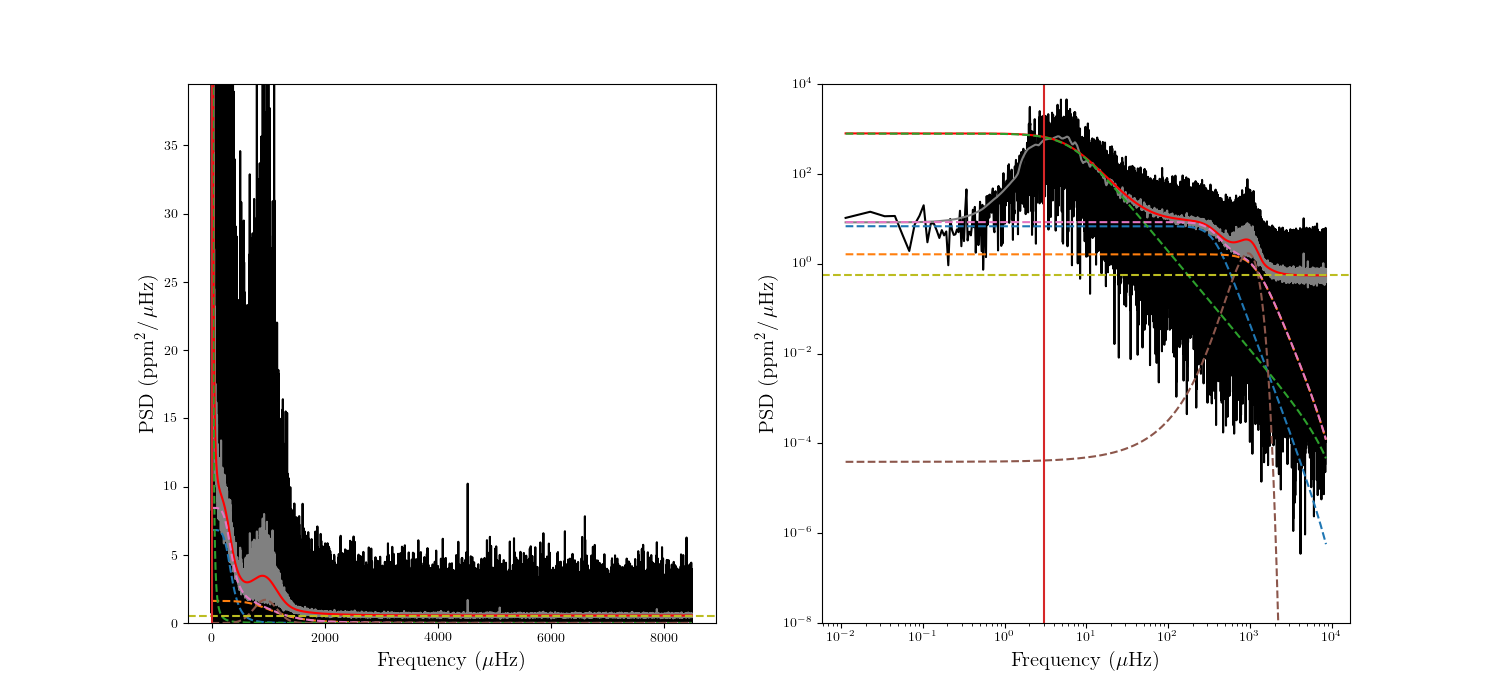}
    \caption{Example of background model to the power spectra of KIC 6679371. A similar approach was applied to all the selected targets of the LEGACY sample. The blue and orange dashed fits correspond to the first and second granulation component, while the pink dashed line is the sum of the the two granulation components. The green dashed fit represents the activity contribution, the yellow dashed line is the white noise level, and the brown fit corresponds to the p-mode contribution. The total model is the solid red fit.}
    \label{fig:bgmodel-example}
\end{figure*}

Here we show the components of the background model to the power spectra of the selected sample of LEGACY stars, as was explained in Section~\ref{sec:RMSKepler}. Fig.~\ref{fig:bgmodel-example} shows the different components of the background model for the star KIC 6679371.

\bsp	
\label{lastpage}
\end{document}